\documentclass{aa}  
\usepackage{graphicx}
\usepackage{txfonts}
\usepackage{lipsum}
\usepackage{subfig}
\usepackage{lscape}                                          
\usepackage{placeins}            
\usepackage{float}
\usepackage{placeins}
\usepackage{amsmath}
\usepackage{multirow}
\usepackage[colorlinks=true, allcolors=black]{hyperref}

\begin{document}

   \title{Magnetic geometry of M dwarfs in the southern PLATO field}
   \titlerunning{Magnetic geometry of M dwarfs in the southern PLATO field}


\author{
M. Diez\inst{1},
P. I. Cristofari\inst{2},
J. Morin\inst{1},
P. Petit\inst{3},
S. Bellotti\inst{2,3},
A. Vidotto\inst{2},
A. Carmona\inst{3},
X. M. Delfosse\inst{4},
C. P. Folsom\inst{5},
G. A. J. Hussain\inst{6},
A. F. Lanza\inst{7} and
S. Messina\inst{7}
}

\authorrunning{Diez, M. et al.}

\institute{
Laboratoire Univers et Particules de Montpellier, Université de Montpellier, CNRS, LUPM/UMR 5299, 34095 Montpellier, France\\
\email{manon.diez@umontpellier.fr}
\and Leiden Observatory, Leiden University, PO Box 9513, 2300 RA Leiden, The Netherlands
\and Institut de Recherche en Astrophysique et Planétologie, Université de Toulouse, CNRS, IRAP/UMR 5277, 14 Avenue Edouard Belin, 31400 Toulouse, France
\and Univ. Grenoble Alpes, CNRS, IPAG, 38000 Grenoble, France
\and Tartu Observatory, University of Tartu, Observatooriumi 1, Tõravere 61602, Estonia
\and Science Division, Directorate of Science, European Space Research and Technology Centre (ESA/ESTEC), Keplerlaan 1, 2201 AZ Noordwijk, The Netherlands
\and INAF-Catania Astrophysical Observatory, Via S. Sofia 78, I-95127 Catania, Italy
}

\date{Received 16 March 2026, accepted 06 April 2026}

  \abstract
   {M dwarfs are the most abundant stars in the Galaxy and exhibit diverse magnetic behaviours. While understanding their large-scale magnetic fields is essential for investigating stellar dynamos and assessing the impact of magnetic activity on planetary environments, their magnetic properties and long-term variability remain poorly characterised.}
   {Our aim was to characterise the large-scale magnetic fields of six M dwarfs in the southern \textit{PLATO} field, with rotation periods ranging from approximately 1 to 17 days and masses between 0.26 and 0.64 M$_\odot$. Five of these stars are partially convective, while one is fully convective. These targets extend the mass–rotation diagram into previously unsampled regions.} 
   {We analysed \textit{TESS} light curves to determine accurate rotation periods and optimise phase coverage for our spectropolarimetric observations. SPIRou data were reduced to obtain least-squares deconvolution (LSD) profiles and longitudinal field measurements, while synthetic spectra fitting yielded small-scale field strengths. We then applied ZDI to reconstruct the large‑scale magnetic topologies of the six targets.}
   {We report a wide diversity of magnetic topologies among the six M dwarfs, with three main results: (1) Rapidly rotating ($P_\mathrm{rot} < 2$~d) early M dwarfs can generate dipole-dominated magnetic fields of moderate intensity, similar to those of less massive mid-M dwarfs; (2) Rapidly rotating mid-M dwarfs can generate non-axisymmetric large-scale magnetic fields featuring a significant toroidal component; (3) We report a moderately rotating ($P_\mathrm{rot}\sim 17$~d) early M dwarf featuring a surprisingly weak large-scale magnetic field.}
   {Our findings further highlight the diversity of magnetic field configurations among M dwarfs, including in previously unexplored regions of parameter space. Long-term monitoring of our sample is crucial in order to distinguish persistent features from variability-driven excursions and to characterise the long-term evolution of their surface magnetic fields. Complementary \textit{PLATO} photometry, including flare and spot‑induced variability analyses, will be essential to link surface activity with magnetic properties.}


   \maketitle
   \nolinenumbers


\section{Introduction}

Low-mass stars, and in particular M dwarfs, represent the most abundant stellar component in the solar neighbourhood \citep{2021A&A...650A.201R}. Their low luminosity, combined with an exceptional longevity, makes them favourable natural laboratories for the study of stellar physics and the evolution of life-supporting environments. M dwarfs play a key role in the evolution of planetary systems. Their characteristics, especially in terms of convection and internal structure, allow us to test models that go beyond the framework of more massive stars such as the Sun.

In M dwarfs, magnetic fields are generated by dynamo processes that can differ from the solar case. This difference is particularly pronounced in fully convective M dwarfs, where the absence of a tachocline leads to a fundamentally non-solar dynamo \citep{ 2006A&A...446.1027C, 2015ApJ...813L..31Y}. In partially convective M dwarfs, the role of the tachocline remains debated, and their dynamo mechanisms may also deviate from the solar model \citep{2020ApJ...893..107B}. These dynamos give rise to intense magnetic activity, which manifests at the stellar surface as spots, plages, and flares, and influences both the atmospheric structure and the stellar wind. Such processes have a direct impact on the habitability of planets orbiting these stars by modulating ultraviolet radiation and energetic particle fluxes \citep{2020IJAsB..19..136A, 2020LRSP...17....4C}. In particular, the strong magnetic pressure of active M dwarfs can significantly compress planetary magnetospheres, potentially exposing part of the atmosphere to erosion \citep{2013A&A...557A..67V}. Understanding magnetic activity is therefore essential, both to characterise stellar dynamos and to assess the environmental conditions of planetary systems. Moreover, the magnetic activity of M dwarfs strongly depends on their rotation. Observations \citep{2014ApJS..211...24M,2017ApJ...834...85N} show that these stars maintain high rotation rates longer than solar-type stars before spinning down. M dwarfs maintain their activity at a high level and directly influence the long-term environment of their planets.

Recent observations indicate that M dwarfs can produce coronal mass ejections (CMEs), which are major drivers of space weather and can strongly affect the atmospheres of nearby planets. Direct evidence has been obtained from radio observations of an early M dwarf, where bursts with properties similar to solar Type II events demonstrate that CME plasma can escape the stellar magnetosphere \citep{2025Natur.647..603C}. These detections provide the first observational constraints on the occurrence rate of stellar CMEs and suggest that planets in the close-in habitable zones of M dwarfs may be exposed to frequent and intense CME impacts. In addition to flares \citep{2020AJ....159...60G}, multi-wavelength observations show that M-dwarf activity arises from strong and complex magnetic fields capable of releasing large amounts of energy, driving plasma motions, and forming localised magnetic structures, such as those associated with starspots \citep{2025SciA...11.6116Z,2025ApJ...990L..32L}. Together, these results highlight the diversity and intensity of magnetic phenomena in M dwarfs and emphasise the need for direct measurements of their surface magnetic fields. Complementary photometric analyses of ultracool dwarfs (M4–L4) from \textit{TESS} light curves \citep{2024MNRAS.527.8290P} determined rotation periods and flare events, showing that rotation- and flare-driven activity is more frequent in intermediate M dwarfs (M4–M6) than in cooler objects. These results provide valuable context for spectropolarimetric observations and help to better understand magnetic variability in the low-mass stars.

Zeeman--Doppler imaging (ZDI), enabled by spectropolarimetry, is a powerful tool to reconstruct the topology of surface magnetic fields from polarised spectral line signatures \citep{1989A&A...225..456S, 1997MNRAS.291..658D, 2016ASSL..439..223M}. Zeeman--Doppler imaging has revealed a wide diversity of magnetic configurations among low-mass stars, ranging from strongly dipolar and organised fields to more complex multi-polar structures, reflecting the diversity of stellar dynamos at work \citep{2010MNRAS.407.2269M, 2013A&A...549L...5G}. However, spectropolarimetry, and hence ZDI, primarily probes the large-scale component of the magnetic field. Measurements derived from Zeeman broadening in unpolarised spectra (Stokes\,I) provide access to the average small-scale magnetic field, including fine-structured regions that are invisible to ZDI \citep{2022A&A...662A..41R, 2025A&A...702A.111C}. The combination of Stokes\,I (unpolarised) and Stokes\,V (polarised) measurements thus provides a more complete view of stellar magnetism, linking the large- and small-scale field components to the observed stellar activity \citep{2015ApJ...813L..31Y}. Campaigns carried out with the SPIRou instrument have led to major advances in the understanding of the magnetic activity of M dwarfs. The observed samples mainly include slowly rotating, weakly active stars (rotation periods > 30 days) \citep{2023MNRAS.525.2015D, 2024MNRAS.527.4330L}, but also more active stars (rotation periods < 5 days, \citealt{2024A&A...686A..66B}). Some locations in the parameter space remain poorly covered, notably rapidly rotating early M dwarfs ($<2$~day), and intermediate periods ($\sim$5-20 days), which are of particular interest for extending our current understanding of stellar dynamo regimes.

Magnetic cycles provide crucial insights into stellar dynamos. The solar dynamo operates through differential rotation and convective motions in a partially convective star to generate a well-characterised 22-year magnetic cycle and an associated 11-year activity cycle, as evidenced by sunspots, plages, and chromospheric activity indices. Many M dwarfs are fully convective and lack a tachocline, so their dynamos likely operate via different mechanisms. Long-term monitoring of chromospheric and coronal emission, as well as photometric variability linked to spots and active regions, reveal a complex picture: some M dwarfs exhibit multiple or irregular cycles, while others show no clear periodicity \citep{2016A&A...595A..12S,2016ApJ...830L..27R, 2023A&A...675A.168M, 2025A&A...696A.230I}. 
Spectropolarimetric observations of several M dwarfs now cover more than a decade, revealing various types of long-term variability \citep{2023A&A...676A..56B, 2024A&A...686A..66B, 2025A&A...704A.298B, 2026A&A...707A.397S}. However, only one direct observation of a polarity reversal is reported \citep{2024MNRAS.527.4330L} and no complete magnetic cycle has ever been inferred from spectropolarimetric observations of M dwarfs to date. Therefore, no clear connection could be established between long-term evolution of the activity and surface magnetic field of M dwarfs yet. This situation contrasts with that of their more massive counterparts: solar-type stars. Long-term spectropolarimetric campaigns have begun to reveal that there is a clear connection between the modulation of stellar activity and the underlying magnetic cycles and dynamo mechanisms \citep{2018A&A...620L..11B, 2025arXiv250920202B}. Moreover, theoretical studies are beginning to provide a coherent picture of these cyclic magnetic fields \citep[e.g.][]{2019A&A...623A..54K,2022ApJ...926...21B}.

The PLAnetary Transits and Oscillations of stars (\textit{PLATO}) satellite \citep{2025ExA....59...26R} will provide  multi-year, high-precision photometric monitoring of numerous bright cool stars, presenting a unique opportunity to study activity cycles. The light curves delivered by space-borne photometers  record precious signatures of stellar magnetic activity, including the photometric imprint of the transit of spots and faculae \citep{2019ApJS..244...21S}, of their slow relative drift under the effect of surface shears \citep{2015A&A...583A..65R}, as well as short-term events such as flares and superflares recorded on stars of all activity levels \citep{2020AJ....159...60G}. This rich information will be an indispensable complement to spectropolarimetric campaigns.

In this context, the present study focuses on signatures of magnetic activity in a selection of M dwarfs located in the southern \textit{PLATO} field. The study combines photometric data, notably from \textit{TESS}, with spectropolarimetric observations. We have two main objectives: first, to characterise the presence and duration of magnetic cycles; and second, to investigate their connection with fundamental stellar properties such as rotation and internal structure. We specifically target five very rapidly rotating M dwarfs ($\sim$1 day), as well as one star with a rotation period of $\sim$17 days, exploring regimes that have been poorly covered in previous studies. Among these targets, five stars are partially convective, while one is fully convective, thus providing valuable constraints on how internal structure affects dynamo processes. As well as contributing to the mass–rotation diagram, this work prepares  for the exploitation of future \textit{PLATO} observations, which will provide precise measurements of rotation periods and long-term photometric variability for a much larger sample of M dwarfs. These data will be crucial for mapping magnetic activity in greater detail, improving our understanding of stellar dynamos, and assessing the impact of this activity on the environments of orbiting planets.


\section{Target stars and rotation periods}
\subsection{Sample selection}

As part of a long-term monitoring program aimed at characterising the magnetic activity of nearby M dwarfs, we selected a sample of six targets suited for regular spectropolarimetric observations with SPIRou \citep{2020MNRAS.498.5684D}, the high-resolution near-infrared spectropolarimeter installed on the Canada-France-Hawaii Telescope (CFHT), located atop Maunakea in Hawaii. SPIRou covers the 0.98--2.35\,µm wavelength range and was specifically designed to detect stellar magnetic fields and search for exoplanets around cool stars. It delivers high-resolution polarised spectra and enables long-term radial velocity monitoring. The first observations used in this study were obtained in the second semester of 2024 (2024B), marking the start of an ongoing, long-term monitoring campaign of the southern \textit{PLATO} field (LOPS2) \citep{2025A&A...694A.313N}. This article presents the initial results of this campaign, which aims to regularly observe these M dwarfs, ideally on an annual basis, over several years, in order to track the evolution of their magnetic fields and potential cycles. Complementary observations of M dwarfs in the northern \textit{PLATO} field (LOPN1) are also being conducted using NeoNarval, located atop the Télescope Bernard Lyot (TBL) at the Pic du Midi observatory in France. This allows us to extend the sample size and compare stellar properties across both \textit{PLATO} fields.

Targets were selected within the LOPS2, based on astrometric, photometric, and visibility criteria ensuring both their accessibility from the CFHT and their relevance for high-cadence temporal monitoring. The initial list was compiled from the Gaia EDR3 catalogue \citep{2021A&A...649A...1G} by applying the colour and absolute magnitude selection criteria proposed by \citet{2021A&A...653A..98M}, which are optimised to identify M dwarfs in the \textit{PLATO} Input Catalogue (PIC). A distance limit of 60\,pc was imposed to ensure high-quality astrometric and photometric data. All retained sources are flagged in the PIC, confirming their suitability for combined stellar and planetary studies, and their spectral classification was independently verified by cross-matching with the catalogue of nearby M dwarfs by \citet{10.1093/mnras/stt1436}.

\begin{table*}
\centering
\setlength{\tabcolsep}{3pt}
\caption{Fundamental parameters and rotation periods of the six M dwarf targets observed during semester 2024.}
\small
\label{tab:sample_prot_transposed}
\begin{tabular}{lcccccc}
\hline\hline
Name & AP Col & CD--35 2213 & CD--26 4156 & CD--35 2722 & CD--29 4446 & PM J05408--3323 \\
\hline
Spectral Type (SpT) & M5 & M4 & M1 & M1 & M1 & M2 \\
Mass ($M_\odot$)$^a$ & $0.263\pm0.022$ & $0.401\pm0.021$ & $0.546\pm0.021$ & $0.552\pm0.021$ & $0.637\pm0.021$ & $0.597\pm0.021$ \\
$T_\mathrm{eff}$ [K]$^b$ & $3035\pm30$ & $3291\pm30$ & $3606\pm30$ & $3707\pm30$ & $3651\pm30$ & $3783\pm30$ \\
log(g) [dex]$^b$ & $4.37\pm0.05$ & $4.79\pm0.05$ & $4.57\pm0.05$ & $4.61\pm0.05$ & $4.71\pm0.05$ & $4.75\pm0.05$ \\
\text{[M/H]} [dex]$^b$ & $-0.01\pm0.10$ & $0.12\pm0.10$ & $0.25\pm0.10$ & $0.14\pm0.10$ & $0.21\pm0.10$ & $0.06\pm0.10$ \\
Inclination $i_\mathrm{zdi}$$^c$ & 60$^\circ$ & 60$^\circ$ & 60$^\circ$ & 45$^\circ$ & 60$^\circ$ & 60$^\circ$ \\
$v \sin(i)$ [km/s]$^b$ & $15.94\pm0.09$ & $20.27\pm0.15$ & $21.77\pm0.13$ & $12.42\pm0.04$ & $21.56\pm0.31$ & $3.70\pm0.05$ \\
$\log(L_X/L_\mathrm{bol})$$^d$ & -- & -- & $-3.22$ & $-3.25$ & $-2.96$ & $-4.40$ \\
$P_\mathrm{rot}$ (lit.) [d] & $1.01^e$ & $1.93\pm0.009^f$ & $1.33^g$ & $1.71^f$ & $1.64\pm0.090^h$ & -- \\
$P_\mathrm{ TESS}$ [d]$^i$ & $1.013 \pm 0.011$ & $1.937 \pm 0.057$ & $1.326 \pm 0.020$ & $1.719 \pm 0.030$ & $1.642 \pm 0.056$ & $17.0 \pm 2.6$ \\
$P_\mathrm{ZDI}$ [d]$^j$ & $0.9940 \pm 0.0009$ & $1.9441 \pm 0.0072$ & $1.3287 \pm 0.0014$ & $1.7220 \pm 0.0050$ & $1.6357 \pm 0.0031$ & $16.5 \pm 1.4$ \\
\hline
\end{tabular}
\tablefoot{
$^a$ Stellar masses estimated using the mass-luminosity-metallicity relation provided by \citet{2019ApJ...871...63M}. \\
$^b$ $T_{\mathrm{eff}}$, $\log (g)$, [M/H], and $v\sin (i)$ derived from the fitting of synthetic spectra \citep{2023MNRAS.522.1342C,2023MNRAS.526.5648C}. The uncertainties correspond to the formal errors from the posterior distributions and do not include systematic sources of error.
 \\
$^c$ The inclination angle used for ZDI was deduced from the stellar radius, $v \sin(i)$ 
 and the rotation period ($P_\mathrm{ZDI}$).
 \\
$^d$ X-ray to bolometric luminosity ratios ($\log(L_X/L_\mathrm{bol})$)  from \citet{2022A&A...661A..29M}. \\
Literature rotation periods compiled from: e) \citet{2023AJ....166...16P}, 
f) \citet{2020ApJ...895..140H}, 
g) \citet{2022A&A...661A..29M}, 
h) \citet{2021A&A...645A..30Z}. When uncertainties are not listed, this indicates they were either not calculated by the authors or not provided in the original source. \\
$^i$ Stellar rotation periods obtained in the present study from \textit{TESS} light curves. \\
$^j$ Stellar rotation periods $P_\mathrm{ZDI}$ obtained from ZDI analyses (basis for ephemeris calculation).
}
\end{table*}

Further constraints were then applied to ensure the feasibility and efficiency of SPIRou spectropolarimetric observations. First, only stars with declination greater than $-38^\circ$ were considered, ensuring a minimum visibility window of two hours per night at airmass $<$\,2 from Maunakea. This corresponds to the northern part of the LOPS2 field, accessible from the northern hemisphere and overlapping the southern continuous viewing zone (CVZ) of \textit{TESS} \citep{2025A&A...694A.313N}.

Second, we selected targets with H-band magnitude $<$\,8.3. This threshold ensures a signal-to-noise ratio (S/N) of approximately 200 that can be reached within 30 minutes per polarimetric sequence, including overheads. This ensures that each polarimetric sequence achieves a sufficient signal-to-noise ratio within a reasonable telescope time, enabling regular long-term monitoring of stellar rotation and magnetic variability.

Each target’s neighborhood was verified using the Gaia EDR3 and 2MASS (Two Micron All Sky Survey; \citealt{2003tmc..book.....C}) catalogues. We excluded all stars with neighbouring sources within the 15$\arcsec$ \textit{PLATO} or 21$\arcsec$ \textit{TESS} pixel sizes, as photometric measurements integrate flux over large angular scales and are therefore sensitive to contamination. This criterion is more restrictive than that required for SPIRou observations, whose $1.3\arcsec$-diameter input fiber is automatically free of contamination once the photometric constraint is satisfied. This ensures that the observed variability in the light curves and the spectropolarimetric signals can be reliably attributed to the target star.

An additional selection criterion was based on stellar rotation. Only stars with either a known rotation period in the literature or a clear periodic signal in the \textit{TESS} light curve were retained. All selected stars exhibit measurable flux rotational modulation. A systematic re-evaluation of the rotation periods was conducted as part of this work, and is described in detail in Section~\ref{subsec:RotationPeriod}. Rotation is a key parameter for understanding stellar magnetism, and accurate knowledge of the period is essential for phase-resolved observations and for characterising large-scale magnetic field geometries.

The final sample includes six M dwarfs, mainly early-type (M1). Later spectral types (up to M5) are less represented in the sample due to the magnitude limit. The stellar masses range between 0.26 and 0.64~$M_\odot$. Effective temperatures were estimated from Gaia colours using the calibration of \citet{2021A&A...653A..98M}, while stellar masses were derived from absolute K-band magnitudes using the Mass-Luminosity-Metallicity relation of \citet{2019ApJ...871...63M}. Estimates based on synthetic spectral fits were also obtained, showing good agreement with values derived from Gaia colours and K-band magnitudes.

Two targets present specific characteristics. CD--35\,2722 is known to host a long-period substellar companion detected through astrometry \citep{wahhaj_2011}, while V372\,Pup (CD--29\,4446, GJ\,2060A) is a double-lined spectroscopic binary (SB2) with well-constrained orbital parameters \citep{2023A&A...672A..82L}. Despite the challenges posed by their multiplicity, both targets were retained in our sample, as their properties are well characterised and their companions are distant enough to allow reliable analysis, with tailored tools available to disentangle their spectroscopic signatures \citep{2024A&A...682A..77T}.

Finally, the sample characterisation was complemented with additional parameters from the literature, when available. These include the projected rotational velocity ($v\sin i$) as a measure of stellar rotation, as well as coronal activity indicators such as X-ray luminosity, compiled from sources such as \citet{2012AcA....62...67K} and \citet{2022A&A...661A..29M}. A summary of the fundamental parameters for the six targets is provided in Table \ref{tab:sample_prot_transposed}.


\subsection{Rotation periods}
\label{subsec:RotationPeriod}

Accurate knowledge of stellar rotation periods is essential to optimise the scheduling and phase coverage of spectropolarimetric observations \citep{2023MNRAS.525.2015D, 2023A&A...672A..52F}. To this end, we use photometric time series primarily from the \textit{TESS} \citep{2015JATIS...1a4003R} mission. The combination of their long-term coverage and data quality enables the reliable detection of periodic modulations induced by stellar activity.

The data processing is performed on all available sectors for each target, without any prior selection. For our six targets, this corresponds to three to six sectors each. For some targets, the same sector is available through multiple \textit{TESS} data products, either due to different exposure times or because the target was re-observed in later cycles. In such cases, we retained only one observation per sector to avoid redundancy. This approach ensures optimal use of the temporal coverage, thereby avoiding biases introduced by partial or sector-limited analyses. The exclusion of certain biased temporal windows, notably related to instrumental effects, is conducted based on the \texttt{SAP\_QUALITY} flags in the \textit{TESS} files, following \citet{2024A&A...687A.208P}. This step automatically removes corrupted or questionable measurements. Data processing is carried out using the Python module \texttt{Lightkurve} \citep{2018ascl.soft12013L}, which provides robust tools for importing, cleaning, and analysing light curves. 

Frequency analysis is performed by computing Lomb–Scargle \citep{1976Ap&SS..39..447L,1982ApJ...263..835S,2018ApJS..236...16V} periodograms on the assembled light curves, enabling the identification of dominant rotation-related periods. Continuous portions of the light curve between gaps larger than 0.1 days were treated as individual segments. Within each segment, long-term trends were removed using a quadratic polynomial fit, effectively flattening the flux while preserving rotational signals. The detrended segments were then concatenated directly to form a single time series spanning all available sectors for each target. This approach ensures that the frequency analysis captures the true periodic modulation due to stellar rotation rather than slow instrumental or astrophysical trends. This method has been validated on a sample larger than the six main targets, including several dozen stars with diverse stellar properties.

\begin{figure}[!htbp]
    \centering
    \includegraphics[width=0.83\columnwidth]{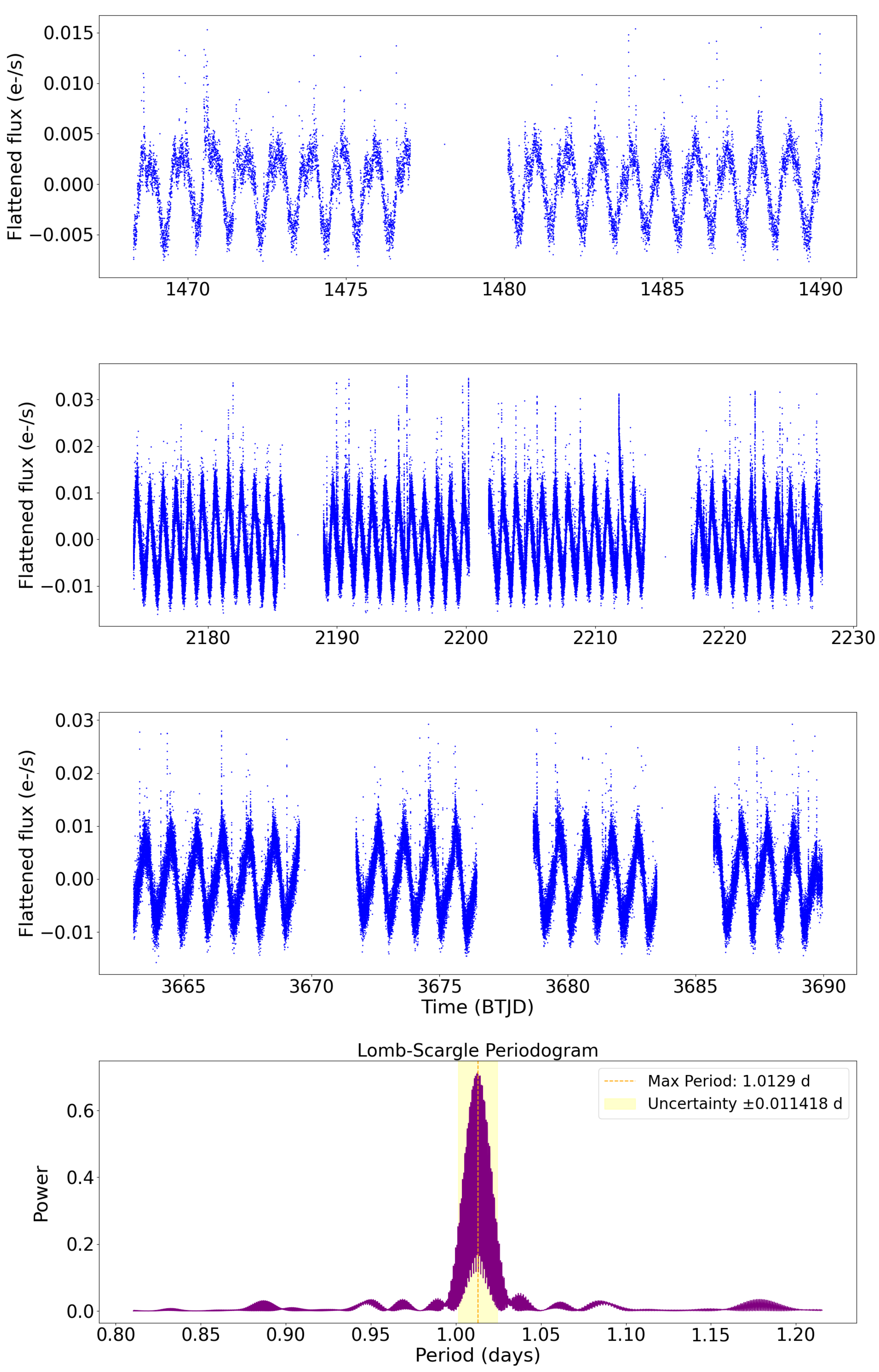}
    \caption{\small Light curves and Lomb-Scargle periodogram for the star AP Col observed by \textit{TESS}. The first, second, and third panels show light curves for four sectors. The second panel combines two consecutive sectors observed one after the other. The fourth panel displays the Lomb-Scargle periodogram computed from all sectors combined, with an estimation of the period uncertainty indicated by the yellow shaded area.}
    \label{fig:APColrot}
\end{figure}

The rotation periods derived from the \textit{TESS} light curves for our six targets range from about 1 to 17~days. These photometric periods were used to plan the spectropolarimetric monitoring campaigns and to ensure adequate phase coverage during observations. Although the final rotation periods adopted in this work ($P_{\mathrm{ZDI}}$) were subsequently refined through Zeeman--Doppler imaging (see Section~4), the photometric determinations are overall consistent with these values, confirming the reliability of our initial estimates. Additionally, Figure \ref{fig:APColrot} presents a concrete example for the target AP Col, showing the combined light curve over all analysed sectors as well as its Lomb-Scargle periodogram, clearly highlighting the detection of the dominant rotation period (see Appendix \ref{AppendixA} for the other five targets).

We note that the detrending can sometimes remove part of the rotationally modulated signal. This can lead to the preferential detection of harmonics, typically at half the true rotation period ($P_{\mathrm{rot}}/2$), instead of the fundamental period.
This effect becomes significant for rotation periods around or above ~10–15 days, when the signal varies on a timescale comparable to gaps, discontinuities, or the sector length (27 days). In this context, PM~J05408$-$3323 represents a specific case: its rotation period is substantially longer than those of our other targets, so the quadratic polynomial detrending was not applied. Due to the presence of additional peaks at $\sim$8.06~days and $\sim$2.93~days (see Appendix~\ref{fig:PMJ054083323Prot}), applying this correction would have preferentially highlighted these signals rather than the primary rotation period of $\sim$17~days, effectively removing a large fraction of the rotational modulation. This example highlights the need to adapt the detrending strategy to the temporal scales relevant to each target, as well as to the characteristics of the available data.

\section{Observations}

The spectropolarimetric data obtained with SPIRou under Program ID 24BF030 were reduced using \textit{A PipelinE to Reduce Observations} (APERO), a reduction package installed at CFHT \citep{2022PASP..134k4509C}. In this study, we performed two complementary analyses. First, we analysed polarised spectra to derive measurements of the longitudinal magnetic field. Second, we modelled the small-scale magnetic field using synthetic spectra, in order to estimate its surface intensity and to derive robust fundamental stellar parameters. These two approaches are described below.

\subsection{Spectral analysis}

We carried out an analysis of the six M dwarfs in our sample to derive their atmospheric properties: effective temperature ($T_\mathrm{eff}$), surface gravity ($\log (g)$), and metallicity ([M/H]), total unresolved photospheric magnetic field, and projected rotational velocity ($v\sin i$). Our analysis relied on synthetic spectra computed from MARCS stellar atmosphere models \citep{2008A&A...486..951G} using the \texttt{ZeeTurbo} code \citep{2023MNRAS.522.1342C}. Our process relies on the direct comparison of the synthetic spectra to the observed spectra in carefully selected spectral lines (see Table \ref{lines} in Appendix \ref{AppendixB}) relying on the tools used in previous studies \citep{2023MNRAS.526.5648C, 2025A&A...702A.111C}. An example of best fit obtained for CD-29~4446 is presented in Fig.~\ref{plotZeeturbo} in Appendix \ref{AppendixB}.

We built a grid of synthetic spectra computed for $T_{\rm eff}$ ranging from 2700 to 4400~K in steps of 100~K, $\log(g)$ ranging from 3.5 to 5.0~dex in range of 0.5~dex and $\rm [M/H]$ ranging from -0.75 to 0.75~dex in steps of 0.25~dex. For set of atmospheric parameters, models were computed for radial magnetic field strengths of 0 to 10~kG in steps of 2~kG. The model spectra ($S_{\rm tot}$) are obtained as a linear combination of the synthetic spectra computed for different field strengths ($S_i$) so that

\begin{equation}
S_{\rm tot} = \sum a_{i}S_{i} \\,
\end{equation}

where $a_{i}$ are filling factors associated with each magnetic component, with $\sum a_{i}=1$.

Our process relies on the exploration of the parameter space with a Markov chain Monte Carlo algorithm. Estimates of the filling factors and atmospheric parameters are obtained from the posterior distributions. Bayesian uncertainties on atmospheric parameters were increased as in \cite{2022MNRAS.511.1893C}, by quadratically adding 30~K for the temperature, 0.05~dex for the gravity, and 0.10~dex for the metallicity, in order to account for some of the systematic inherent to the models.
For the six targets in our sample, our process was applied to template spectra built by taking the median of our observations, in order to increase the signal-to-noise ratio (S/N).

For our six M dwarfs, we obtain $T_\mathrm{eff}$ ranging from 3035 to 3783~K, $\log (g)$ ranging from 4.37 to 4.79, and $[M/H]$ ranging from -0.01 to 0.25 (Table~\ref{tab:sample_prot_transposed}). Our results suggest that four of the stars in our sample are metal-rich. Small-scale magnetic fields fall between 0.3 and 4.4~kG (Table~\ref{tab:tech}).

\subsection{LSD profile from polarised spectra}
\label{subsec:3.2}

The least-squares deconvolution (LSD) technique was applied to extract a synthetic mean circularly polarised profile by combining several thousand photospheric atomic lines. This method, originally introduced by \citet{1997MNRAS.291..658D}, was implemented in the Python code \texttt{LSDpy},\footnote{\url{https://github.com/folsomcp/LSDpy}} developed by C.P.~Folsom \citep{SpecpolFlow2025}. It significantly enhances the signal-to-noise ratio of the Stokes~$I$ and $V$ profiles by leveraging the redundancy in the spectral information. 

The LSD line masks were constructed using the Vienna Atomic Line Database (VALD)\footnote{\url{http://vald.astro.uu.se/}} \citep{2015PhyS...90e4005R} to provide the atomic line list, combined with a MARCS atmosphere model \citep{2008A&A...486..951G} appropriate for M dwarfs.
 Typical stellar parameters were adopted for the masks (effective temperature, surface gravity, and microturbulence). The mask includes several hundred lines in the infrared, with known Landé factors and depths sufficient to contribute to the LSD signal. Only photospheric lines with reliable atomic data were retained, while broad hydrogen lines, regions affected by telluric absorption, and lines with atypical shapes (extended damping wings, \citealt{2010A&A...524A...5K}) were excluded. Only lines with central depths larger than approximately 5\% of the continuum were retained, in order to discard very weak lines dominated by noise. Each selected line was assigned a weight according to its expected contribution to the signal. For circular polarisation, this depends on line depth, wavelength and effective Landé factor; for intensity, it depends on line depth. The resulting LSD profiles were then scaled to reference values of central wavelength, depth, and Landé factor, producing a representative average line suitable for consistent comparison across observations.

From the resulting LSD profiles, the longitudinal magnetic field $B_z$, was computed using the first-order moment method, following the formalism of \citet{1979A&A....74....1R} as adapted by \citet{1997A&A...326.1135D} and \citet{2000MNRAS.313..851W}. The expression used is

\begin{equation}
B_z \, (\mathrm{G}) = -\frac{2.14 \times 10^{11}}{\lambda_0\, g_{\mathrm{eff}}\, c} \cdot \frac{\int v\, V(v)\, dv}{\int \left(I_c - I(v)\right) dv} \\,
\label{eq:Bz}
\end{equation} where $v$ is the radial velocity in the stellar rest frame (in km\,s$^{-1}$), $\lambda_0$ is the mean wavelength of the LSD profile (in nm), $g_{\mathrm{eff}}$ is the effective Landé factor, $c$ is the speed of light in the same units as $v$, $V(v)$ and $I(v)$ are respectively the LSD Stokes~$V$ and $I$ profiles, and $I_c$ is the continuum level (normalised to 1). For each star, we adopted a normalisation wavelength of 1700\,nm and an effective Landé factor of 1.2144. The integration was performed around the line center over a range of $\pm 42$ to $\pm 58$\,km\,s$^{-1}$ depending on the star, in order to fully encompass the absorption regions of both Stokes $I$ and $V$ profiles \citep{2023A&A...676A..56B}.

\begin{figure}[H]
\centering
\includegraphics[width=\columnwidth]{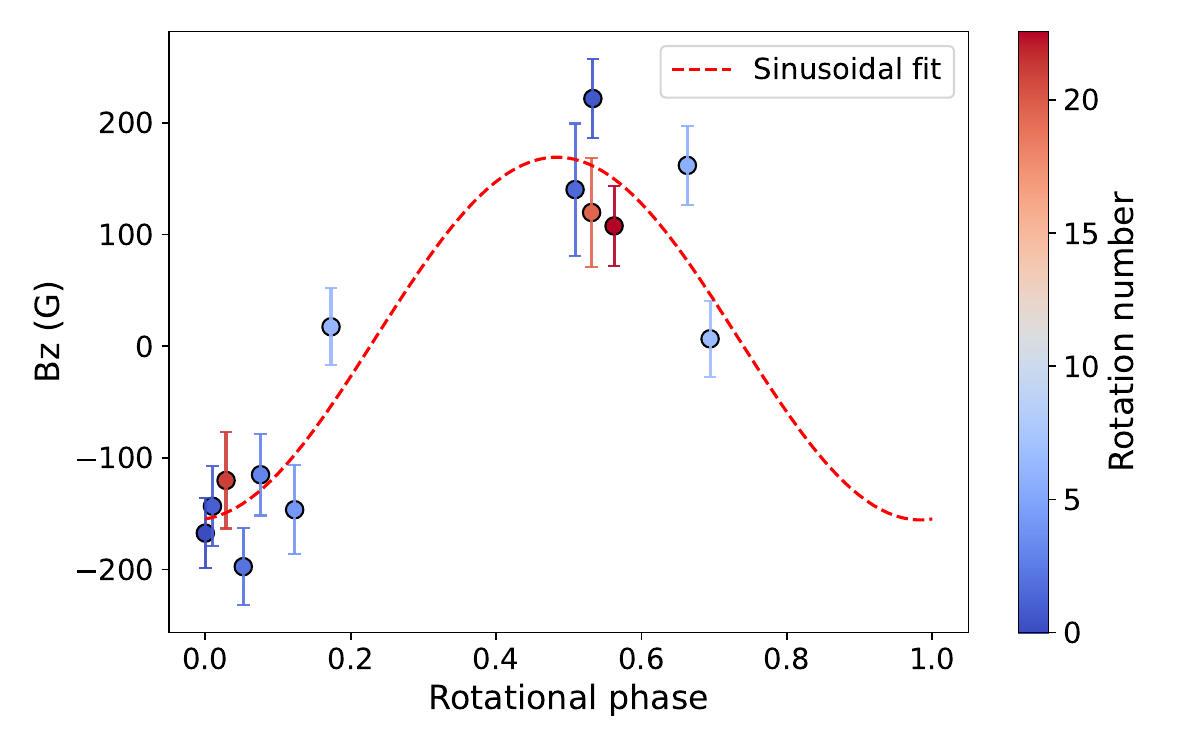}

\vspace{1em}

\includegraphics[width=\columnwidth]{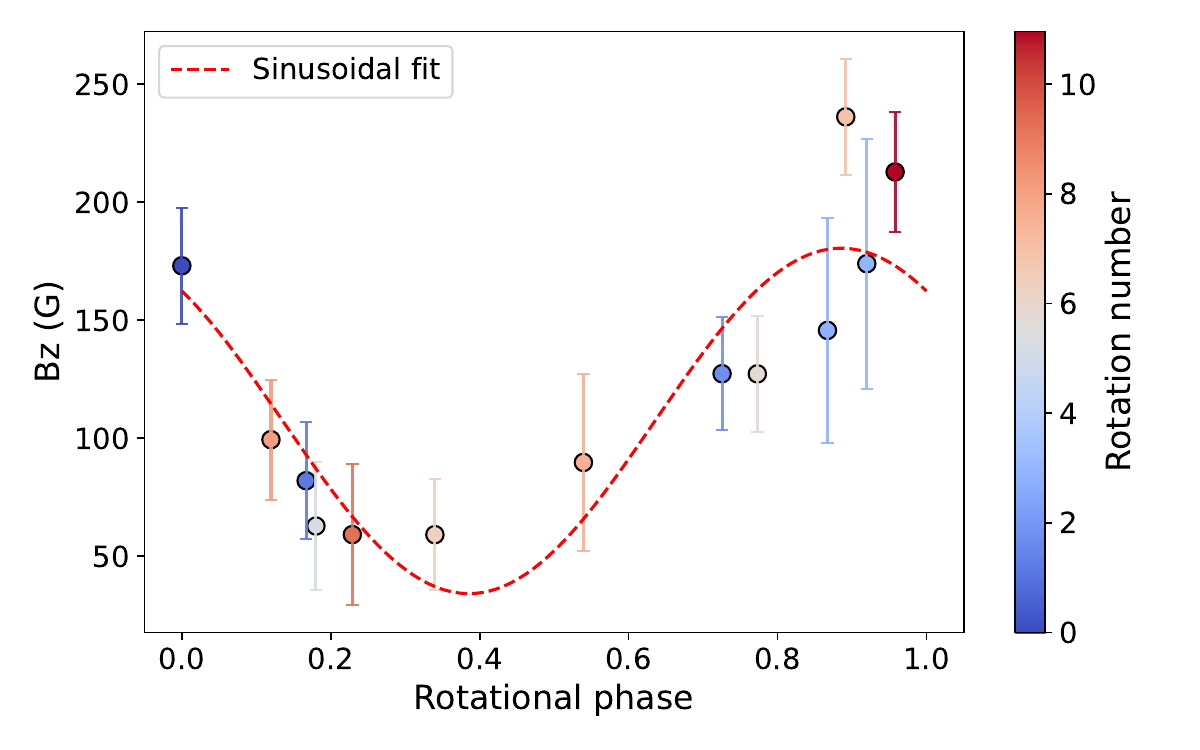}

\caption{\small Longitudinal magnetic field \( B_z \) as a function of rotational phase for the stars CD$-$35\,2213 (top) and CD$-$35\,2722 (bottom). The colour bar indicates the number of rotations.}
\label{fig:Bz-phase_duo}
\end{figure}

This calculation was performed using the \texttt{specpolFlow} Python package\footnote{\url{https://github.com/folsomcp/specpolFlow}} \citep{SpecpolFlow2025}, which also provides associated uncertainties. For the observed targets during the 2024B run, we summarise their main properties and the extent of the spectropolarimetric monitoring. The results are summarised in Table \ref{tab:tech}, which lists, for each target, the number of observations, measured $B_z$ values, and corresponding errors. The temporal evolution of $B_z$ was plotted as a function of rotational phase (see Fig.\ref{fig:Bz-phase_duo}). While the plots presented in this paper use the rotation periods refined via ZDI, the initial rotation periods determined for these stars were used at the time of the observations to assess the phase coverage. This allows us to evaluate how well our observations sampled the rotation cycle. Overall, the measurements are well distributed and coherent. In cases where the coverage is not optimal (for instance for AP Col, owing to a rotation period close to 1 d), points acquired at different rotational phases still provide consistent information. These variations provide valuable constraints on the large-scale field geometry. Examples for CD-35 2213 and CD-35 2722 are shown in Figure~\ref{fig:Bz-phase_duo} (see Appendix~\ref{AppendixC} for the other targets).

\begin{table}[ht]
\centering
\caption{Magnetic characteristics of the stars in our sample.}
\scalebox{0.90}{%
\begin{tabular}{l c c c c c }
\hline
ID 
& $N_{\mathrm{obs}}$
& $\langle |B_z| \rangle$
& $\mathrm{std}(B_z)$
& $\langle u(B_z) \rangle$
& $\langle B_I \rangle$ \\
& 
& {[G]}
& {[G]}
& {[G]}
& {[kG]} \\
\hline
AP~COL          & 19 & 356$\pm$10 & 67  & 56 & 3.1$\pm$0.14  \\
CD-35\,2213     & 13 & 128$\pm$10 & 146 & 39 & 4.4$\pm$0.14  \\
CD-26\,4156     & 14 & 64$\pm$10  & 49  & 32 & 3.2$\pm$0.05  \\
CD-35\,2722     & 13 & 127$\pm$9 & 59  & 30 & 3.5$\pm$0.04 \\
CD--29 4446     & 23 & 64$\pm$7 & 39  & 30 & 3.1$\pm$0.06  \\
PM~J05408-3323  & 12 & 6$\pm$4   & 9   & 10 & 0.3$\pm$0.04  \\
\hline
\end{tabular}
}
\tablefoot{ Summary of observations in 2024B: 1) Target name, 2) Number of observations, 3) Average absolute value of longitudinal magnetic field \( \langle |B_z| \rangle \), 4) Standard deviation of \( B_z \) values, 5) Average uncertainty of \( B_z \), 6) Average small-scale magnetic field strength at the stellar surface \( B_I \).}
\label{tab:tech}
\end{table}

\section{Magnetic Doppler imaging}

Zeeman--Doppler imaging (ZDI; \citealt{1989A&A...225..456S, 1997A&A...326.1135D}) is a tomographic inversion technique used to reconstruct the large-scale magnetic field at the surface of a rotating star, based on a time series of high-resolution circularly polarised spectra. The method exploits the rotational modulation of Zeeman signatures in the Stokes~V profiles, which are sensitive to the line-of-sight component of the magnetic field vector.

We model the stellar surface using a spherical grid and describe the magnetic field as a combination of spherical harmonic modes, including both poloidal and toroidal components \citep{2006MNRAS.370..629D}. We compute the synthetic Stokes I and V profiles using the Unno–Rachkovsky solutions to the polarised radiative transfer equations in a Milne–Eddington atmosphere \citep{1956PASJ....8..108U,1967IzKry..37...56R,2004ASSL..307.....L}. This approximation assumes depth-independent physical parameters and a linear source function, providing a physically consistent and analytical description of the Zeeman effect, which is suitable for Zeeman–Doppler imaging. We also incorporate the filling-factor formalism commonly adopted in ZDI \citep{2008MNRAS.390..567M,2023A&A...676A..56B}, which separates magnetic and non-magnetic contributions to Stokes I and adjusts the fraction of the surface that contributes to the polarised signature in Stokes V. For our dataset, the observed Stokes V signatures can be satisfactorily reproduced without a polarisation filling factor, so we adopt $f_V = 1$ \citep{2008MNRAS.384...77M, 2008MNRAS.390..545D}. Local Stokes profiles computed in this framework are then integrated over the visible stellar disk, taking Doppler shifts associated with stellar rotation and limb darkening into account.
Finally, magnetic inversion is performed using maximum entropy regularisation to select the simplest magnetic topology compatible with the data, mitigating the non-uniqueness of the inverse problem.

We use the ZDIpy ZDI code\footnote{\url{https://github.com/folsomcp/ZDIpy}} developed by \citet{2018MNRAS.474.4956F}, implemented in \texttt{Python}, which simultaneously fits a series of LSD Stokes~V profiles to infer the surface magnetic field configuration. The model is initialised with no magnetic field, so that the initial reduced $\chi^2$ corresponds to a null-field model. Input parameters include stellar inclination, projected rotational velocity ($v \sin i$), rotation period, and local line properties (central wavelength, Landé factor, line depth). The rotation period is taken from ZDI ($P_{\rm ZDI}$, Table \ref{tab:sample_prot_transposed}), $v \sin i$ is measured from Stokes I as described in Section 3.1, and the stellar inclination is derived from $P_{\rm ZDI}$, $v \sin i$, and the stellar radius. When this calculation formally yields inclinations close to or above 90°, we adopt $i = 60^\circ$ following \citet{2008MNRAS.390..567M, 2010MNRAS.407.2269M}, ensuring stable and consistent ZDI reconstructions. For the line models, the wavelength and Landé factor are the same as those used for the calculation of $B_z$ in Section 3.2, namely 1700 nm and 1.2144. We adopt a linear limb-darkening coefficient of 0.3 \citep{2011A&A...529A..75C}. The output is a vector magnetic map that provides information on the geometry and strength of the field, essential for characterising the magnetic activity and underlying dynamo processes in M dwarfs.

Building on the ZDI framework, we investigated the presence of surface differential rotation by exploring a range of rotation periods and differential rotation rates ($d\Omega$) for each target. For each combination, a magnetic map was reconstructed and the fit between observed and synthetic LSD profiles was evaluated through the reduced $\chi^2$, while keeping the map entropy constant to allow a fair comparison, following the maximum entropy approach of \citet{1984MNRAS.211..111S} and its ZDI application by \citet{2002MNRAS.334..374P}. Optimal parameters were identified at the minimum of the $\chi^2$ distribution, and uncertainties were estimated from the $\chi^2$ variations around this minimum \citep{1976ApJ...208..177L, 1976ApJ...210..642A, 1992nrfa.book.....P}. 

In the ZDI reconstructions, the final reduced $\chi^2$ for each target was systematically chosen slightly above the values obtained from the Null profiles. Since the Null $\chi^2_r$ values are very low, this procedure leads to a modest overestimation of the error bars, but it does not affect the overall reliability of the reconstructed magnetic maps.


\subsection{AP Col}

AP~Col is an M5 star, initially identified as a nearby pre-main-sequence star and proposed as a member of the Argus/IC~2391 association \citep{2011AJ....142..104R}, but more recent kinematic and photometric analyses based on Gaia EDR3 data support its membership in the AB~Doradus moving group \citep{2022MNRAS.511.6179L}. This young age (\(\sim 130\text{--}200~\mathrm{Myr}\); \citealt{2015MNRAS.454..593B}) is consistent with the relatively low $\log g$ value we derive compared to the other stars in our sample. AP Col has an estimated mass of 0.263~$M_\odot$ and a rotation period of approximately 0.99~days (see Table~\ref{tab:sample_prot_transposed}). Nineteen observations were obtained between 9 November 2024 and 18 December 2024, covering roughly 39 rotations. Due to the rotation period revised via ZDI $(P_{\mathrm{ZDI}} = 0.9940\,\mathrm{d})$, which is very short and very close to the daily duty cycle of observations, the phase coverage is not optimal. Nevertheless, several points acquired at different rotations remain coherent, indicating a stable magnetic geometry over the timespan of data collection. The initial reduced $\chi^2$ of the Stokes V profiles was 3.74 and the final reduced $\chi^2_r$ was 1.10.

\begin{figure}[htbp]
    \centering
    \includegraphics[width=0.53\hsize]{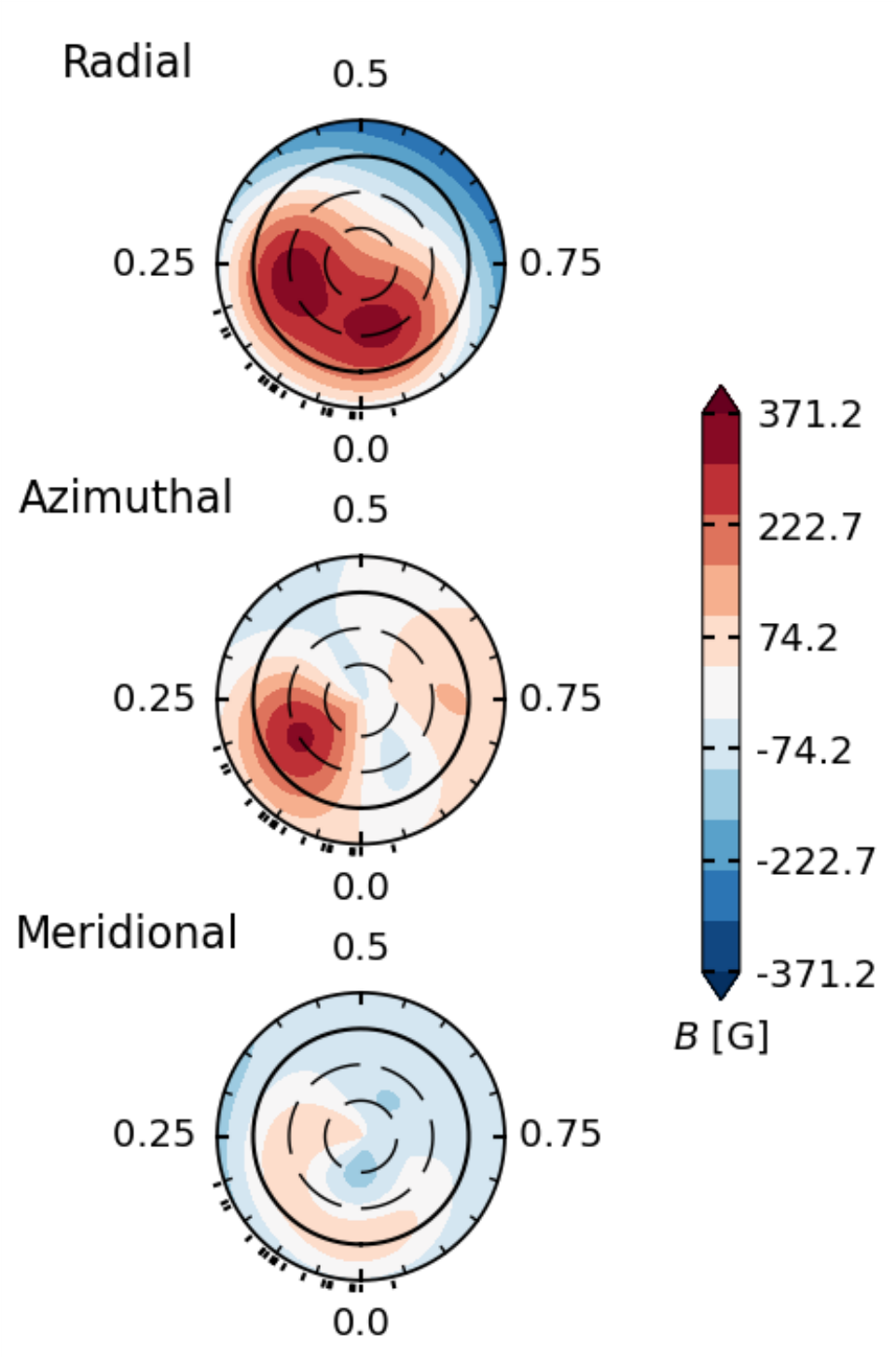}
    \caption{\small Zeeman--Doppler imaging map of the large-scale magnetic field at the surface of AP Col. 
    The radial (top), azimuthal (middle), and meridional (bottom) components of the magnetic field vector are displayed. 
    The colour bar range is set by the maximum of the magnetic field and illustrates the positive (red) and negative (blue) polarity.}
    \label{fig:carte_mag_AP_Col}
\end{figure}

The mean surface field of AP Col is 224~G. AP Col exhibits a large-scale magnetic field dominated by a dipolar component (see Fig.~\ref{fig:carte_mag_AP_Col}). The magnetic field is mainly poloidal and axisymmetric, with the poloidal axisymmetric component representing 60~\% of the poloidal energy. The dipole contributes the majority of the poloidal energy (85~\%). In these initial reconstructions, based on the rotation period derived from \textit{TESS} data (1.0129~d), the field was roughly twice as strong and significantly more axisymmetric. This difference is very likely due to phase coverage, when the observations are folded with the more accurate rotation period, the phase distribution is less homogeneous, weakening the constraints on the ZDI inversion. As discussed by \cite{2008MNRAS.390..567M} in their study of EQ~Peg~B, incomplete phase coverage can bias the reconstruction, particularly by artificially reducing the recovered axisymmetric fraction and large-scale field strength. The solution obtained with the revised period should therefore be interpreted with caution, as the current phase coverage may lead to an underestimation of the magnetic energy and axisymmetry.

Given the high degree of axisymmetry, the search for surface differential rotation did not yield a significant measurement. The reduced $\chi^2$ landscape is essentially flat, indicating that any differential rotation is negligible within the current dataset. Therefore, the rotation period was refined under the assumption of no differential rotation ($d\Omega = 0$), yielding a more precise estimate of $P_{\rm rot} = 0.9940 \pm 0.0009$~days. It should be noted that the $d\Omega = 0$ assumption is based solely on the currently available observations. Future measurements, covering a longer timespan or different epochs, could reveal the presence of surface differential rotation and lead to a revised estimate of $P_{\rm rot}$.


\subsection{CD-35 2213}

CD‑35 2213 is an M4 star with an estimated mass of 0.401~$M_\odot$ and a rotation period of approximately 1.94~days (see Table~\ref{tab:sample_prot_transposed}). CD--35~2213 is a binary system with an orbital period of approximately 2.5~years \citep{Bowler_2015,2014AJ....147...85R}. It is therefore reasonable to conclude that the secondary does not significantly affect our analysis. A comparison between the 2024B observations presented here and additional data obtained in 2025B (to be published separately) shows that the Stokes~V signatures are consistently associated with the primary component, with no detectable contribution from the secondary. We note that the presence of its binary companion affects the stellar characterisation and may lead to an overestimate of $v \sin{i}$. 
\begin{figure}[H]
    \centering
    \includegraphics[width=0.53\hsize]{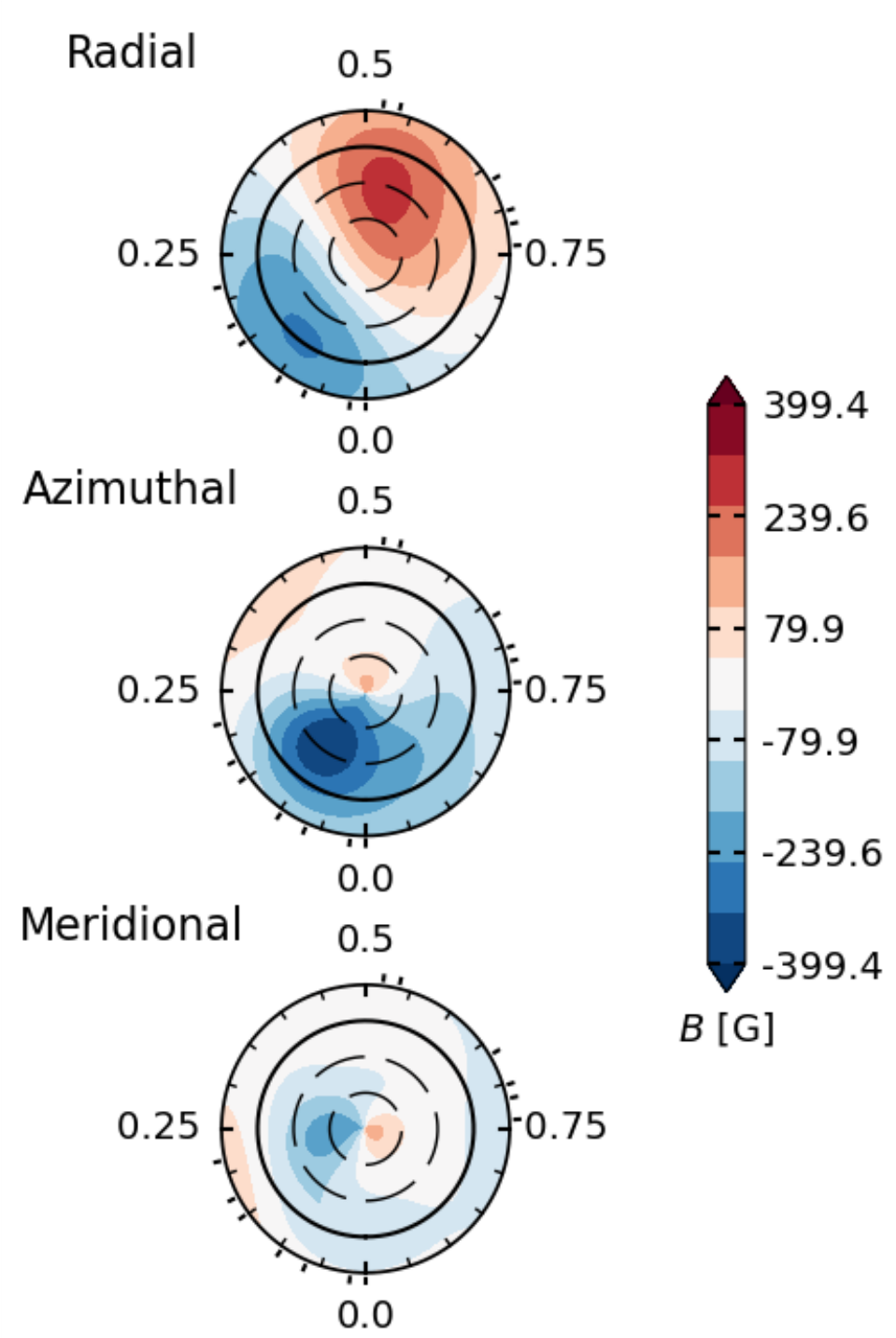}
    \caption{\small Same as Figure \ref{fig:carte_mag_AP_Col}, but for CD-35 2213.}
    \label{fig:carte_mag_CD-35_2213}
\end{figure}

Thirteen observations were obtained between 9 November 2024 and 23 December 2024, covering roughly 23 rotations. While this phase coverage is therefore sub-optimal, spectra recorded over the different rotations indicate that the phase dependence of $B_z$ remains stable. The initial reduced $\chi^2$ of the Stokes V profiles was 1.78 and the final reduced $\chi^2_r$ was 0.90. The mean surface field of CD‑35 2213 is 191~G. CD‑35 2213 exhibits a mainly poloidal magnetic topology, largely dominated by a non-axisymmetric dipolar component, with a significant and mostly axisymmetric toroidal contribution (see Fig.~\ref{fig:carte_mag_CD-35_2213}). The poloidal component represents 79~\% of the total magnetic energy, with the dipole contributing 86~\% of the poloidal energy. The poloidal axisymmetric fraction is 12~\%, while the total toroidal fraction, which is significant, is 21~\%, of which 91~\% is axisymmetric.

A preliminary search for surface differential rotation, limited by the sparse phase coverage of the observations, yields $d\Omega = -0.0263 \pm 0.0632$~rad~d$^{-1}$, with a corresponding equatorial period of $P_\mathrm{eq} = 1.9453 \pm 0.0063$~days for a reduced $\chi^2_r$ of 0.8986. Since $d\Omega$ is consistent with zero within the uncertainty, the rotation period was refined assuming no differential rotation ($d\Omega = 0$), giving $P_\mathrm{eq} = 1.9441 \pm 0.0072$~days.


\subsection{CD-26 4156}

CD‑26 4156 is an M1 star with an estimated mass of 0.546~$M_\odot$ and a rotation period of approximately 1.33~days (see Table~\ref{tab:sample_prot_transposed}). Fourteen observations were obtained between 15 November 2024 and 23 December 2024, covering roughly 28 rotations. The phase coverage could be improved with additional observations; nevertheless, the points acquired remain coherent, indicating a stability of the magnetic geometry. The initial reduced $\chi^2$ of the Stokes V profiles was 1.98 and the final reduced $\chi^2_r$ was 0.95.

\begin{figure}[H]
    \centering
    \includegraphics[width=0.53\hsize]{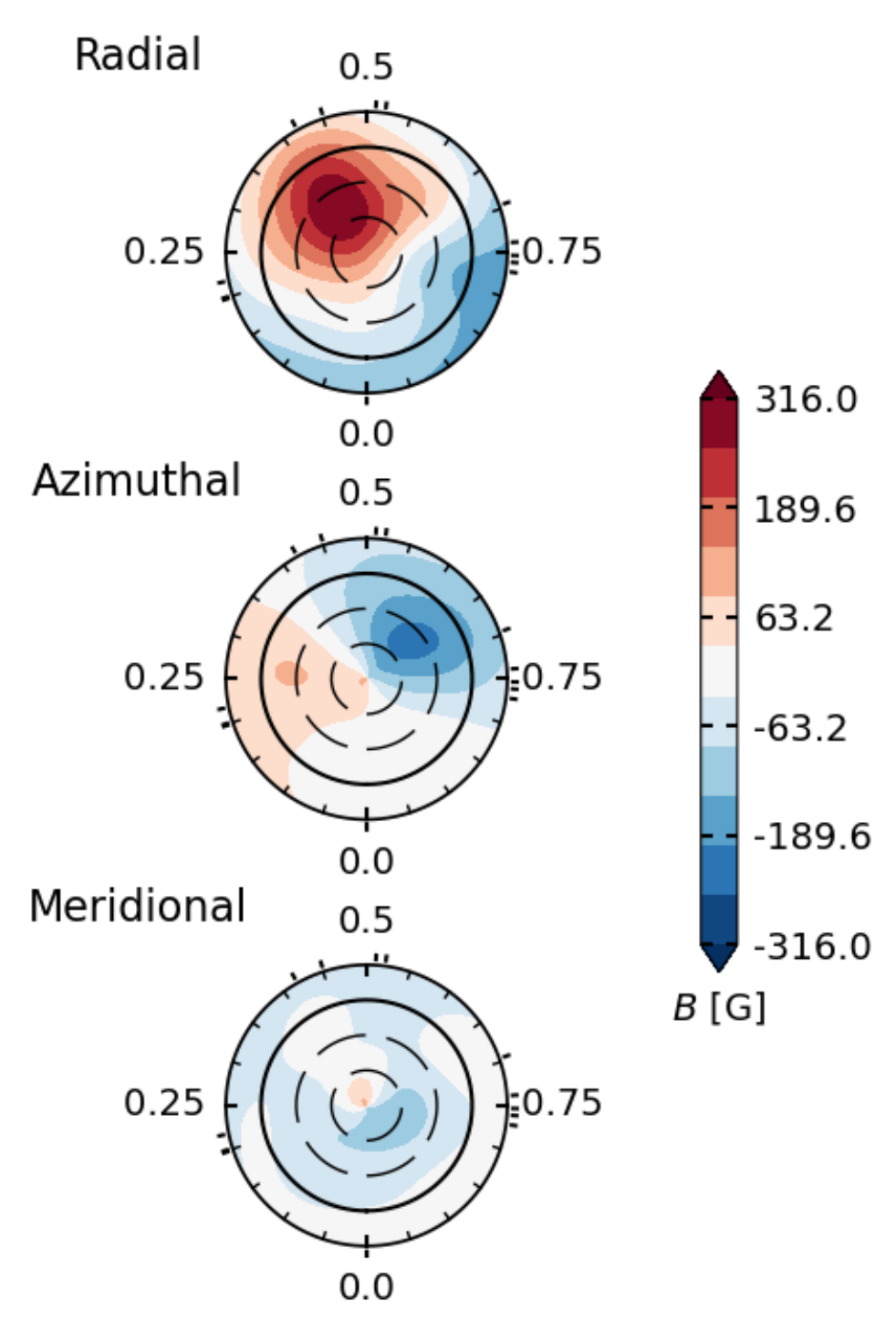}
    \caption{\small Same as Figure \ref{fig:carte_mag_AP_Col}, but for CD-26 4156.}
    \label{fig:carte_mag_CD-26_4156}
\end{figure}

The mean surface field of CD‑26 4156 is 149~G. CD‑26 4156 exhibits a predominantly poloidal magnetic topology, strongly dominated by a dipolar component, with a minor but non-negligible toroidal contribution (Fig.~\ref{fig:carte_mag_CD-26_4156}). The poloidal component represents 95~\% of the total magnetic energy, with the dipole contributing 89~\% of the poloidal energy. The poloidal axisymmetric fraction is 37~\%, while the total toroidal fraction is 5~\%, of which 75~\% is axisymmetric.

A preliminary analysis concerning the search for surface differential rotation indicates $d\Omega = 0.0189 \pm 0.0200$~rad~d$^{-1}$, with a corresponding equatorial period of $P_\mathrm{eq} = 1.3274 \pm 0.0016$~days for a minimum reduced $\chi^2_r$ of 0.8995. Since $d\Omega$ is consistent with zero within the uncertainty, the rotation period was refined assuming no differential rotation ($d\Omega = 0$), giving $P_\mathrm{eq} = 1.3287 \pm 0.0014$~days.


\subsection{CD-35 2722}

CD‑35 2722 is an M1 star with an estimated mass of 0.552~$M_\odot$ and a rotation period of approximately 1.72~days (see Table~\ref{tab:sample_prot_transposed}). Thirteen observations were obtained between 4 and 23 December 2024, covering roughly 11 rotations. The phase coverage is excellent, and the points acquired are coherent, indicating a stability of the magnetic geometry. The star hosts a confirmed substellar companion \citep{2011ApJ...729..139W}, whose atmosphere has been studied \citep{2025A&A...701A..51P}. The initial reduced $\chi^2$ of the Stokes V profiles was 2.19 and the final reduced $\chi^2_r$ was 1.00.

\begin{figure}[H]
    \centering
    \includegraphics[width=0.53\hsize]{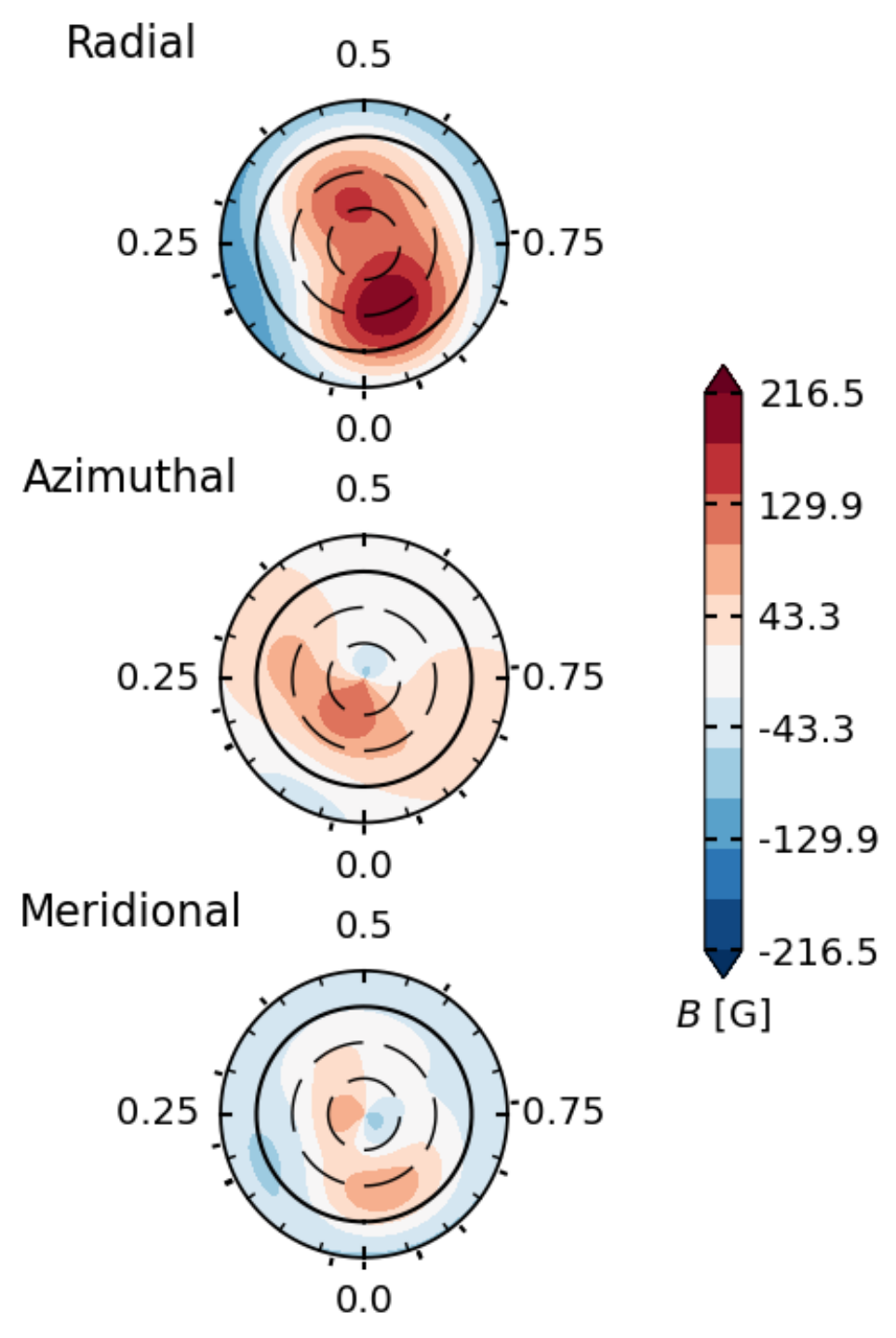}
    \caption{\small Same as Figure \ref{fig:carte_mag_AP_Col}, but for CD-35 2722.}
    \label{fig:carte_mag_CD-35_2722}
\end{figure}

The mean surface field of CD‑35 2722 is 113~G. CD‑35 2722 exhibits a predominantly poloidal magnetic topology, strongly dominated by a dipolar component, with a minor toroidal contribution (Fig.~\ref{fig:carte_mag_CD-35_2722}). The poloidal component represents 93~\% of the total magnetic energy, with the dipole contributing 77~\% of the poloidal energy. The poloidal axisymmetric fraction is 76~\%, while the dipole axisymmetric fraction is 91~\%. The total toroidal fraction is 7~\%, of which 78~\% is axisymmetric.

\begin{table*}
\centering
\caption{Magnetic field characteristics of the six targets observed during semester 2024B.}
\label{tab:magnetic_table3}
\resizebox{\textwidth}{!}{%
\begin{tabular}{lcccccccccccccc}
\hline
Name & Mass & $P_\mathrm{ZDI}$ & $\chi^2_r$ & $\langle |B_v| \rangle$ & $\langle |B_v| \rangle$/$\langle B_I \rangle$ & $f_{\mathrm{pol}}$ & $f_{\mathrm{tor}}$ &
$f_{\mathrm{dip}}$ & $f_{\mathrm{quad}}$ & $f_{\mathrm{oct}}$ &
$f_{\mathrm{axi}}$ & $f_{\mathrm{axi,pol}}$ & $f_{\mathrm{axi,tor}}$ \\
 & [$M_\odot$] & [d] &  & [G] & [\%] & [\%] & [\%] & [\%] & [\%] & [\%] & [\%] & [\%] & [\%] \\
\hline
\rule{0pt}{2.5ex}
AP Col            & $0.263\pm0.022$ & $0.9940 \pm 0.0009$ & 1.10 & 224 & 7 & 94 & 6 & 85 & 11 & 4 & 62 & 60 & 83 \\
CD--35 2213       & $0.401 \pm 0.021$ & $1.9441 \pm 0.0072$& 0.90 & 191 & 4 & 79 & 21 & 86 & 12 & 2  & 29 & 12 & 91 \\
CD--26 4156       & $0.546 \pm 0.021$ & $1.3287 \pm 0.0014$ & 0.95 & 149  & 5 & 95 & 5  & 89 & 7  & 3  & 39 & 37 & 75\\
CD--35 2722       & $0.552 \pm 0.021$ & $1.7220 \pm 0.0050$ & 1.00 & 113 & 3 & 93 & 7  & 77 & 19  & 3  & 76 & 76 & 78 \\
CD--29 4446       & $0.637 \pm 0.021$ & $1.6357 \pm 0.0031$ & 0.90 & 100 & 3 & 86 & 14 & 94 & 2  & 3  & 92 & 91 & 98 \\
PM J05408--3323   & $0.597 \pm 0.021$ & $16.5 \pm 1.4$ & 0.93 & 8 & 3 & 81 & 19 & 93 & 5 & 2 & 64 & 55 & 99 \\
\hline
\end{tabular}%
}
\par\vspace{2mm}
{\footnotesize
\textbf{Notes.} For each target the following quantities are listed in order: star name, stellar mass (references are given in Table~\ref{tab:sample_prot_transposed}), rotation period from ZDI, final reduced $\chi^2_r$ of the Stokes~V fit, mean unsigned magnetic field $\langle |B_v| \rangle$, ratio of the mean unsigned magnetic field to the average small-scale magnetic field at the stellar surface $\langle |B_v| \rangle / B_I$, and the fractional magnetic energies: poloidal ($f_{\mathrm{pol}}$), toroidal ($f_{\mathrm{tor}}$), dipolar ($f_{\mathrm{dip}}$), quadrupolar ($f_{\mathrm{quad}}$), octupolar ($f_{\mathrm{oct}}$), axisymmetric ($f_{\mathrm{axi}}$), poloidal axisymmetric ($f_{\mathrm{axi,pol}}$), and toroidal axisymmetric ($f_{\mathrm{axi,tor}}$).}
\end{table*}

A preliminary analysis concerning the search for surface differential rotation indicates $d\Omega = 0.0211 \pm 0.0726$~rad~d$^{-1}$, with a corresponding equatorial period of $P_\mathrm{eq} = 1.7189 \pm 0.0111$~days for a minimum reduced $\chi^2_r$ of 0.9998. Since $d\Omega$ is consistent with zero within the uncertainty, the rotation period was refined assuming no differential rotation ($d\Omega = 0$), giving $P_\mathrm{eq} = 1.7220 \pm 0.0050$~days.

\subsection{CD-29 4446}

CD‑29 4446 (GJ 2060) is an M1 star of 0.637~$M_\odot$ with a rotation period of approximately 1.64~days (see Table~\ref{tab:sample_prot_transposed}). It is a member of the AB Doradus moving group and forms a spectroscopic binary (SB2) with an orbital period of 7.794 ± 0.008 years \citep{2023A&A...672A..82L}. Twenty-three observations were obtained between 13 October and 23 December 2024, covering roughly 43 rotations. The radial velocity dispersion (interquartile range 0.22~km~s$^{-1}$), due to the SB2 nature of the system, does not affect our analysis. The phase coverage is excellent, and the points acquired are coherent, indicating a stability of the magnetic geometry. The initial reduced $\chi^2$ of the Stokes V profiles was 1.52 and the final reduced $\chi^2_r$ was 0.90.

\begin{figure}[H]
    \centering
    \includegraphics[width=0.53\hsize]{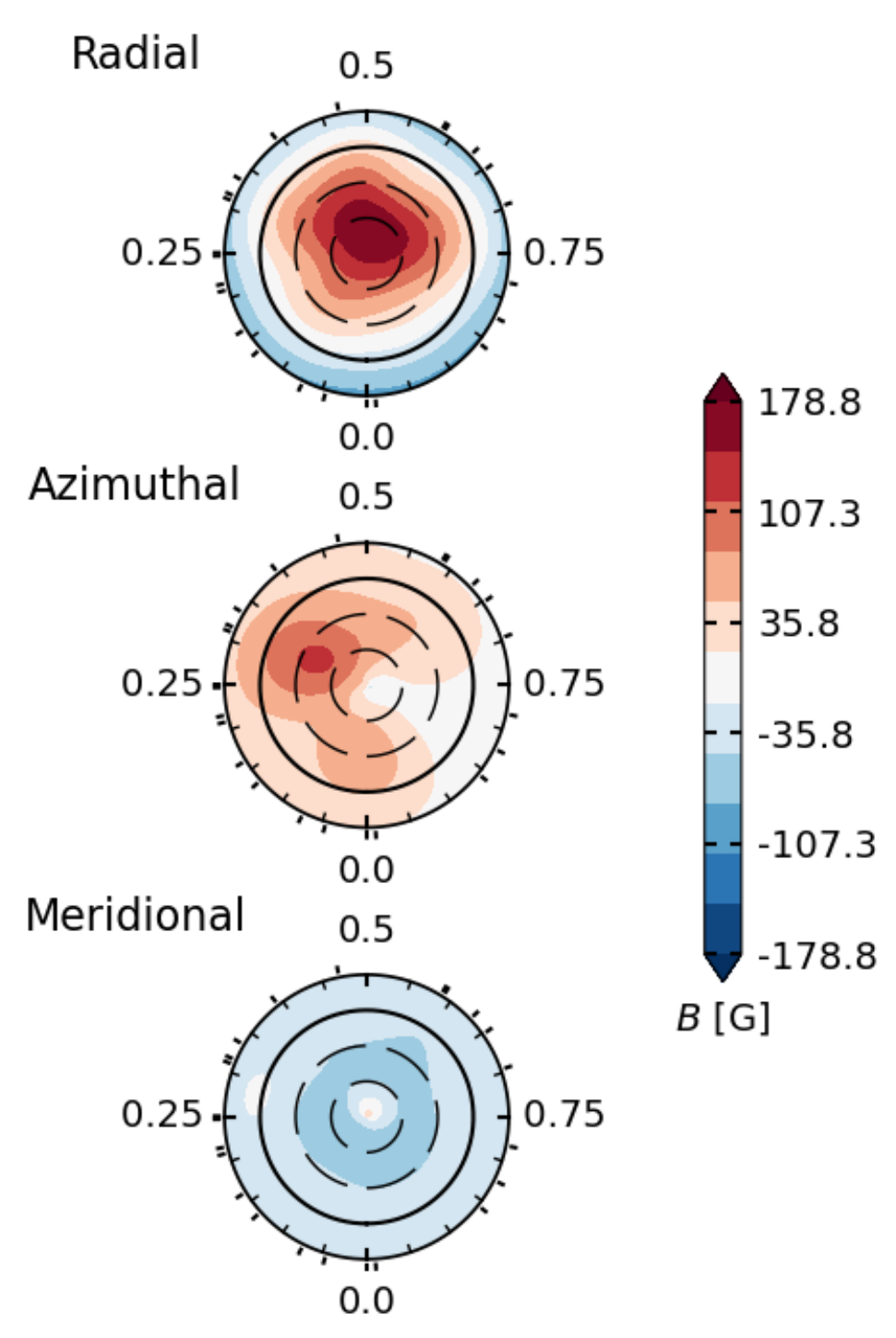}
    \caption{\small Same as Figure \ref{fig:carte_mag_AP_Col}, but for CD-29 4446.}
    \label{fig:carte_mag_CD-29_4446}
\end{figure}

The mean surface field of CD‑29 4446 is 100~G. CD‑29 4446 exhibits a predominantly poloidal magnetic topology, strongly dominated by a dipolar component, with a minor but non-negligible toroidal contribution (Fig.~\ref{fig:carte_mag_CD-29_4446}). The poloidal component represents 86~\% of the total magnetic energy, with the dipole contributing 94~\% of the poloidal energy. The poloidal axisymmetric fraction is 91~\%, while the dipole axisymmetric fraction is 93~\%. The total toroidal fraction is 14~\%, of which 98~\% is axisymmetric.

An analysis of surface differential rotation indicates $d\Omega = -0.0026 \pm 0.0289$~rad~d$^{-1}$, with a corresponding equatorial period of $P_\mathrm{eq} = 1.6358 \pm 0.0055$~days for a minimum reduced $\chi^2_r$ of 0.8976. Since $d\Omega$ is consistent with zero within the uncertainty, the rotation period was refined assuming no differential rotation ($d\Omega = 0$), giving $P_\mathrm{eq} = 1.6357 \pm 0.0031$~days.

\subsection{PM J05408-3323}

PM~J05408$-$3323 is an M2 star with an estimated mass of 0.597~$M_\odot$ and a rotation period of 16.5~days. Twelve observations were collected between 9 November and 23 December 2024, sampling only about three stellar rotations due to the long rotation period of the star. Despite the limited number of cycles, the observations are well distributed in phase and ensure an average uncertainty on $B_z$ as low as 10~G. The weak magnetic signal of the star also requires longer exposure times, contributing to the reduced number of measurements and this relatively low value (compared to other targets investigated here) indicates that the recorded Zeeman signatures were close to our detection threshold. Unlike the other targets in the sample, PM~J05408$-$3323 shows no rotational modulation of $B_z$ (see Appendix~\ref{AppendixC}). The initial reduced $\chi^2$ of the Stokes V profiles was 1.09 and the final reduced $\chi^2_r$ was 0.93. 

\begin{figure}[H]
    \centering
    \includegraphics[width=0.51\hsize]{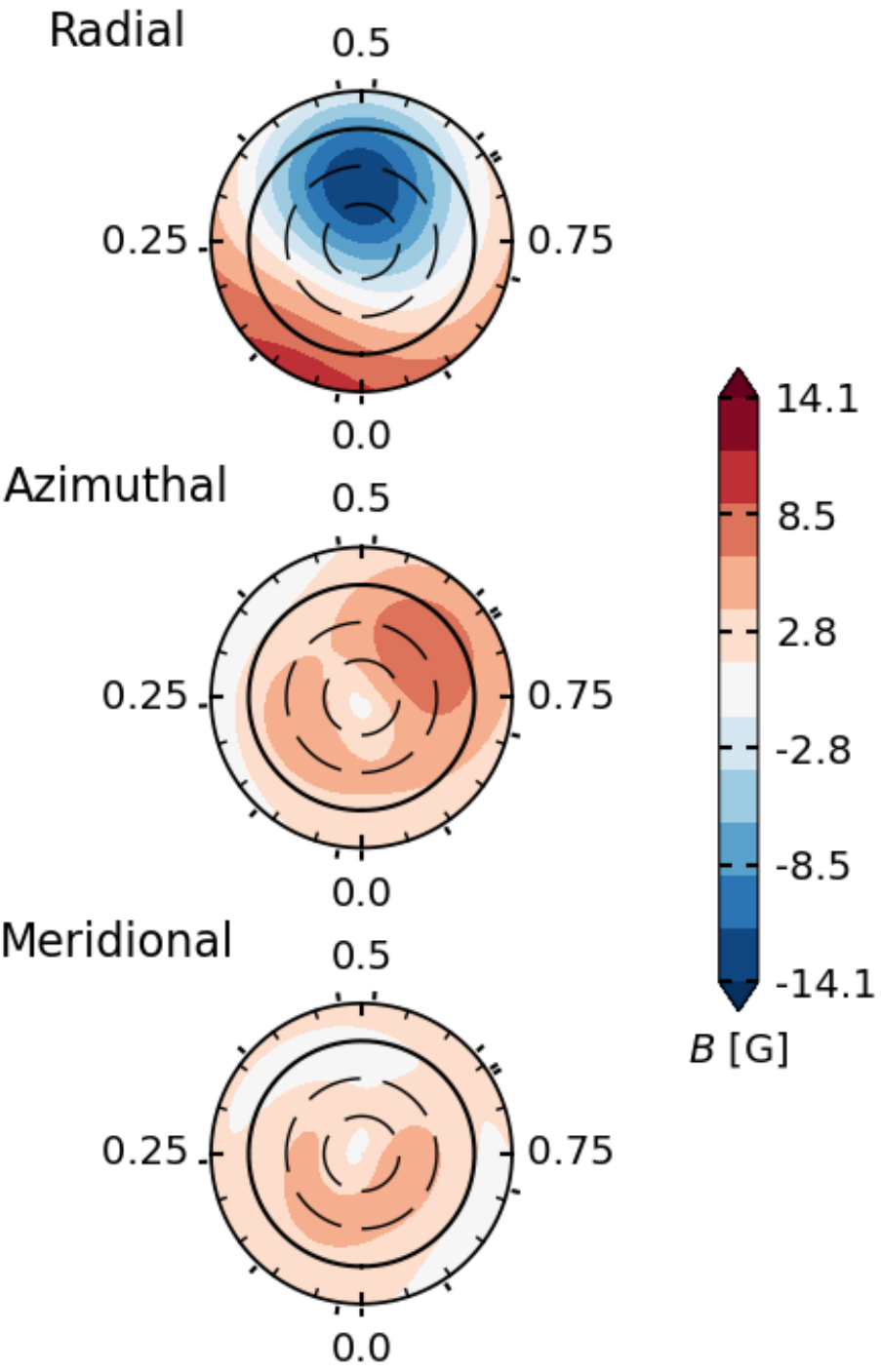}
    \caption{\small Same as Figure \ref{fig:carte_mag_AP_Col}, but for PM J05408-3323.}
    \label{fig:carte_mag_PM_J05408-3323}
\end{figure}

The star hosts a mixed magnetic topology, with a significant poloidal component (81\%) and a non-negligible toroidal contribution (19\%) (Fig.~\ref{fig:carte_mag_PM_J05408-3323}). The poloidal field is strongly dominated by the dipolar mode, which contains 93\% of the poloidal energy. The field geometry shows moderate axisymmetry, with the axisymmetric fraction reaching 55\% of the poloidal component and as much as 99\% of the toroidal field. The mean surface field is very weak, reaching only 8~G, consistent with the small amplitude of the reconstructed magnetic signatures.

A preliminary analysis searching for surface differential rotation yields $d\Omega = -0.0172 \pm 0.1000$~rad~d$^{-1}$, corresponding to an equatorial rotation period of $P_\mathrm{eq} = 16.7 \pm 1.8$~days at a minimum reduced $\chi^2_r = 0.9299$ (initial $\chi^2_r = 1.09$). Given the large uncertainties and the compatibility of $d\Omega$ with zero, the rotation period was recomputed assuming solid-body rotation ($d\Omega = 0$), yielding $P_\mathrm{eq} = 16.5 \pm 1.4$~days. The broad uncertainties reflect both the weak magnetic signal and the limited rotational coverage, making any detection of differential rotation inconclusive.


\section{Discussion and conclusions}


\subsection{Photometry} 

From the photometric perspective, our study of rotation periods presents some inherent limitations linked to the \textit{TESS} data and the methods employed, meaning that this approach cannot be universally applied to any target. A limited number of sectors, poor light curve quality, or the relatively short observing window per sector (typically $\sim$27~days; \citealt{2020ApJS..250...20C}) can make the reliable extraction of rotation periods challenging, especially for slow rotators or for periods near or exceeding $\sim$30~days. Furthermore, the applied detrending can sometimes remove part of the astrophysical signal, leading to the detection of a harmonic rather than the fundamental period, particularly for long periods or when the light curve contains gaps.  

Our six main targets have relatively short rotation periods and high-quality light curves, which greatly reduces these risks and allows a reliable period determination with the available data. Despite the demonstrated effectiveness of our method, these limitations must be considered when planning observations for other targets.   

For instance, among these six targets, PM~J05408$-$3323 exhibits a secondary periodicity around 8.06~days, likely associated with spots on opposite hemispheres, as well as a third notable periodicity at 2.93~days (see Appendix~\ref{fig:PMJ054083323Prot}). The latter may correspond to a previously undetected companion with a rotation period of $\sim$2.9~days;  the high  renormalised unit weight error (RUWE) value reported by Gaia~DR3 \citep{2023A&A...674A...1G} (RUWE~$=$~10) further supports the binary hypothesis \citep{2024A&A...688A...1C} . This example illustrates how additional signals can complicate period extraction and interpretation even in high-quality data.

Significant work also remains to be carried out on the light curves already available, particularly for the analysis of flares, spot-induced variability, and, more broadly, the activity of our targets. Most of the targets in our study exhibit prominent flare activity, as shown by the light curves (see Figure~\ref{fig:APColrot} and Appendix~\ref{AppendixA}). These diagnostics will be essential for linking photometric activity to magnetic topology and for establishing robust correlations between surface variability and magnetic properties in M dwarfs. Finally, future observations from the \textit{PLATO} mission \citep{2025ExA....59...26R}, with longer observing durations and higher-precision photometry, will overcome these constraints and allow  a more detailed investigation of the connection between photometric variability (spots, flares) and magnetic topology.


\subsection{Small-scale magnetic fields}

\begin{figure*}[!htbp]
    \centering
    \includegraphics[width=0.48\textwidth]{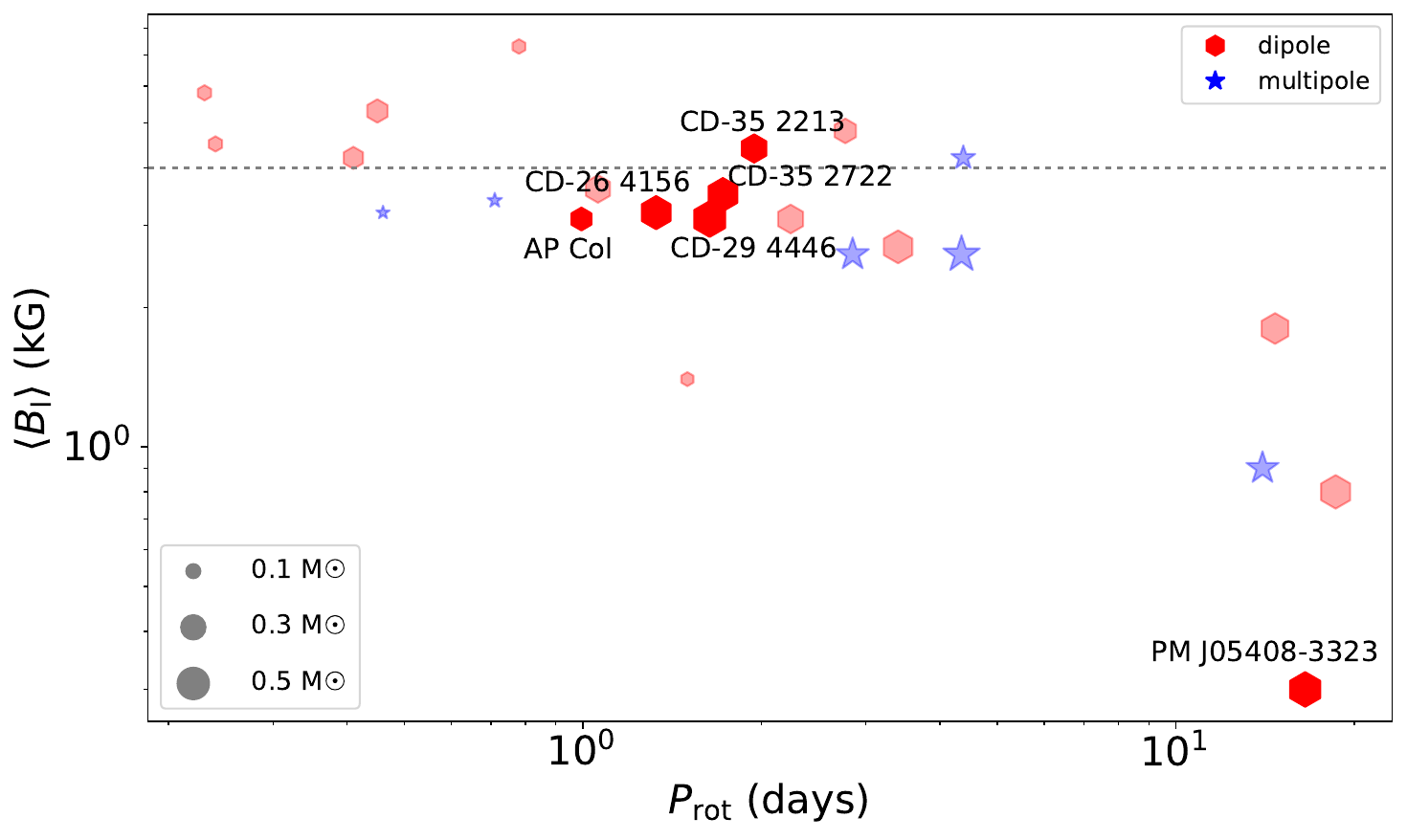}
    \includegraphics[width=0.48\textwidth]{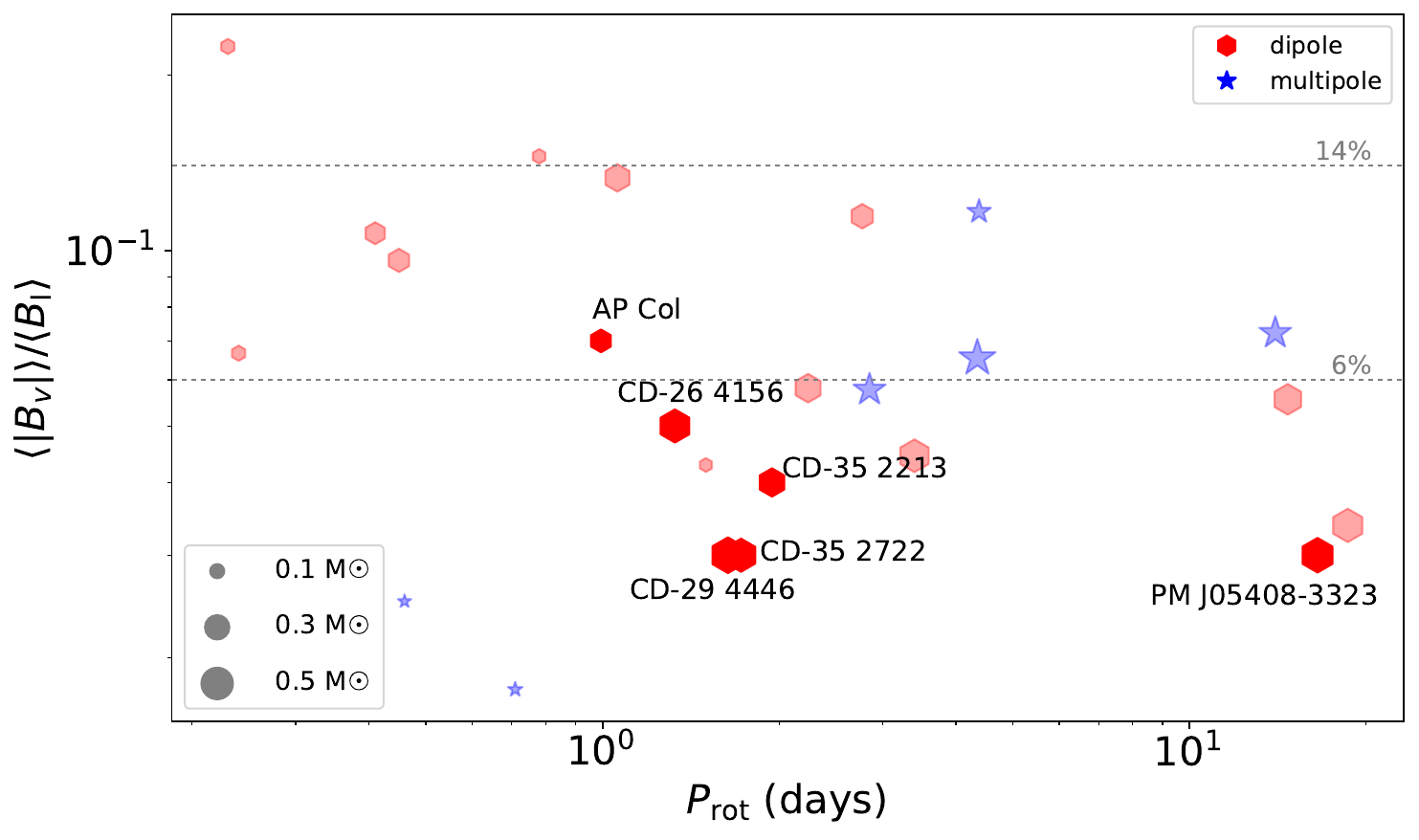}
    \caption{Left: Small-scale magnetic field of M dwarfs as a function of rotation period. Red hexagons and blue stars indicate literature \citep{2010MNRAS.407.2269M,2017NatAs...1E.184S} measurements for stars in dipole and multi-pole states, respectively. The six M dwarfs in this work are shown with the same symbols, fully opaque, with their names indicated on the plot. The symbol size scales with stellar mass (see scale), and the horizontal dashed line shows the 4 kG saturation threshold associated with stars in a multi-pole dynamo regime. Right: Ratio of the magnetic fluxes recovered from Stokes~V to Stokes~I measurements as a function of rotation period. The same symbols and colour-coding are used as in the left panel, and the horizontal dashed lines indicate the mean fraction of magnetic flux detected in Stokes~V spectra (6\% and 14\%).
}
    \label{fig:figure9}
\end{figure*}

Figure~\ref{fig:figure9} presents the magnetic properties of our targets as a function of the rotation period. The left panel shows the small-scale magnetic field strength $\langle B_I \rangle$, while the right panel displays the ratio of the large-scale magnetic field reconstructed by ZDI to the small-scale field ($\langle |B_{\mathrm{v}}| \rangle / \langle B_I \rangle$). We compare our results to those of M dwarfs from the literature, in particular objects studied by \cite{2010MNRAS.407.2269M} and \cite{2017NatAs...1E.184S}, which serve as comparison samples.

The measured values of $\langle B_I \rangle$ for our targets are generally consistent with those reported by \cite{2017NatAs...1E.184S}, who showed that stars hosting dominant dipolar fields—particularly rapid rotators—can reach or even exceed the magnetic saturation level while exhibiting significant intrinsic dispersion (see Figure~\ref{fig:figure9}, left). In our sample, three targets lie slightly below the saturation line, but remain compatible with this dispersion. PM~J05408--3323 stands out with a notably lower $B_I$, while reflecting its slow rotation and weak magnetic field; this measurement is also more uncertain than for the other stars and should therefore be interpreted with caution, placing it in the expected non-saturated regime.  

The ratio $\langle |B_{\mathrm{v}}| \rangle / \langle B_I \rangle$, ranging from $3\%$ to $7\%$, indicates that only a small fraction of the total magnetic field is organised on large spatial scales and detectable in Stokes~V (see Figure~\ref{fig:figure9}, right). This is consistent with previous studies that reported values close to 5~\% for active partly convective M dwarfs and greater than 10~\% for those located just below the fully convective boundary \citep{2009IAUS..259..339R, 2010MNRAS.407.2269M}. In our sample, only AP~Col is fully convective. Its position in Figure~\ref{fig:figure9} (right) is likely affected by an observational bias due to limited phase coverage, caused by its rotation period being very close to 1~d (see Section~\ref{subsec:3.2}), which could lead to an underestimation of $\langle |B_{\mathrm{v}}| \rangle$. PM J05408-3323 is close to the detection limit both in Stokes~I and V, resulting in an uncertain value that may explain its particularly low ratio, and therefore its location in the figure. For the four remaining partially convective stars, the uniformly low $\langle |B_{\mathrm{v}}| \rangle / \langle B_I \rangle$ ratios are consistent with this trend, extending the observations to rapidly rotating partially convective stars.


\subsection{Large-scale magnetic fields}

\begin{figure*}[!htbp]
\centering
\includegraphics[width=\textwidth]{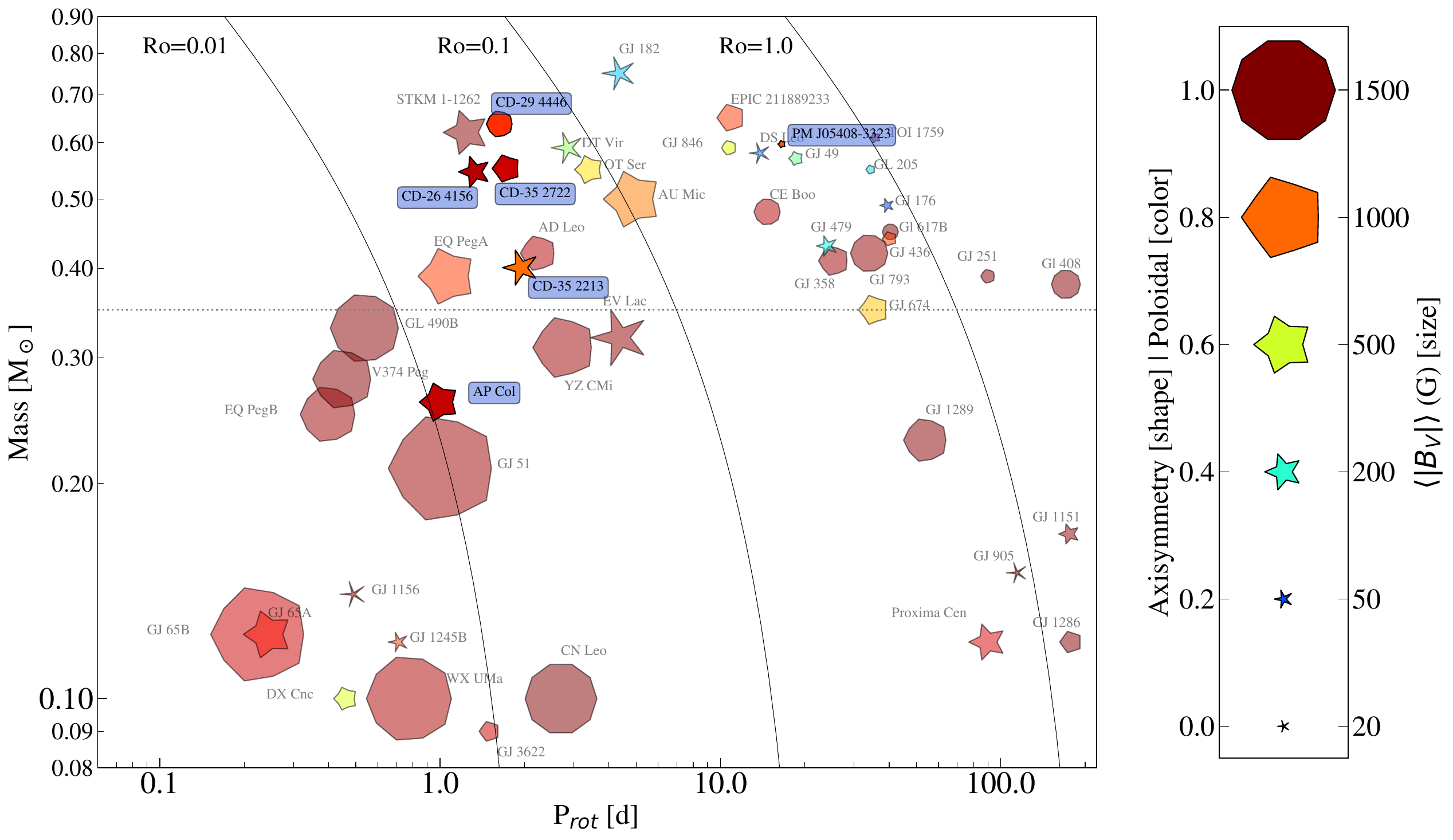}  
\caption{Magnetic properties of cool, single, main-sequence stars, derived from ZDI. Close binaries are excluded, as interactions between the stellar components may affect their rotation and magnetic properties \citep{2024A&A...682A..77T}. The axes represent the stellar rotation period (horizontal) and stellar mass (vertical). The grey curves indicate different Rossby numbers ($Ro = 0.01$, 0.1, and 1.0), calculated using the empirical relations from \citet{2018MNRAS.479.2351W}, which quantify the ratio of rotation to convective turnover time. The symbols in the plot encode three properties of the stellar magnetic field. The size of each symbol indicates the average magnetic field strength; the  larger symbols correspond  to stronger fields. The colour represents the poloidal/toroidal energy fraction: the more the colour shifts towards red, the more poloidal the field; conversely, the more it shifts towards blue, the more toroidal the field. Finally, the shape of the symbol reflects the degree of axisymmetry: the more hexagonal the symbol, the higher the axisymmetry, while more star-like symbols with thinner branches indicate lower axisymmetry (see \citet{2008MNRAS.390..567M}; \citet{2008MNRAS.390..545D}; \citet{2009ApJ...704.1721P}; \citet{2010MNRAS.407.2269M}; \citet{2016MNRAS.461.1465H}; \citet{2017ApJ...835L...4K}; \citet{2017MNRAS.472.4563M}; \citet{2021MNRAS.500.1844K,2021MNRAS.502..188K};  \citet{2022A&A...659A..71W}; \citet{2022A&A...660A..86M}; \citet{2023A&A...676A.139B}; \citet{2024MNRAS.527.4330L}; \citet{2025A&A...704A.298B}). Our six targets, AP Col, CD-35 2213, CD-26 4156, CD-35 2722, CD-29 4446, and PM J05408-3323, are highlighted in blue.}
\label{fig:Confusogram}
\end{figure*}

The inclusion of our six M dwarfs in the mass–rotation diagram (Fig.~\ref{fig:Confusogram}) extends the coverage of regions that were previously underrepresented. Three of our stars, CD–29~4446, CD–26~4156, and CD–35~2722, occupy a region around 1.3–1.7 days of rotation period, masses of 0.55–0.64~M$_\odot$, and Rossby numbers near 0.1, which was previously sparsely sampled. Only StKM~1–1262, which was recently studied by \citet{2025A&A...704A.298B} with similar methods,  occupies a nearby position in the diagram. All four stars share overall similar properties: a reconstructed large-scale magnetic field of moderate intensity $\langle |B_v| \rangle \sim 100-300$~G, mostly poloidal (although CD–29~4446 and StKM~1–1262 display a significant toroidal component of $\sim 14~\%$) with a dominant dipolar component. However, the degree of axisymmetry of the poloidal component  greatly varies from star to star, from $37~\%$ (CD–26~4156) to $91~\%$ (CD–29~4446). As opposed to previously observed M dwarfs of similar mass (DT~Vir and OT~Ser), these four stars clearly belong to the saturated regime of the rotation--activity relation. Our results suggest that their large-scale magnetic field are similar to those of less massive M dwarfs close to the fully convective boundary hosting strong dipole-dominated large-scale fields, such as AD~Leo and EQ~Peg~A. Further monitoring of these targets is required to assess their year-to-year variability and confirm these preliminary conclusions.

CD--35~2213, located near AD~Leo in the diagram, is particularly interesting. The properties of the large magnetic field of this star  differ from those of other rapidly rotating M dwarfs located close to the fully convective boundary. The field of CD–35~2213 is  weakly axisymmetric (29\%), and has a relatively strong toroidal component (21\%), atypical for a star of this mass and rotation period. However, the moderately strong large-scale field $\langle |B_v| \rangle \sim 200$~G and the dipole-dominated poloidal component are in line with the properties of AD~Leo. Regular temporal monitoring is required to determine whether this represents a peculiar excursion in the magnetic properties of CD–35~2213 or a configuration that remains stable over time.

AP~Col, located near V374~Peg, EQ~Peg~B, and GJ~51 in the mass–rotation period diagram (Fig.~\ref{fig:Confusogram}), illustrates the dipole-dominated magnetic configuration of a rapidly rotating low-mass (\(\sim0.3~M_\odot\)) M dwarf.  However, it displays a weaker ($\langle |B_v| \rangle \sim 200$~G) and less axisymmetric (62~\%) large-scale field than the  other three stars. These differences could, at least partly, be a result of the less-than-optimal phase coverage provided by our observations (section~\ref{subsec:3.2}). New observations with more homogeneous phase sampling will be required to more robustly constrain the field geometry and assess its temporal stability.

PM~J05408--3323, with a Rossby number close to unity, occupies an intermediate zone of the diagram near DS~Leo, characterised by weaker fields and more variable topologies. Its field is weak (8~G), predominantly poloidal (81\%), and moderately axisymmetric (64\%). It is the least magnetic star in this region of the diagram, making it particularly noteworthy. While these properties broadly follow the trends identified by \citet{2016MNRAS.462.4442S} for more massive, partially convective stars, stars in this Rossby range can exhibit significant temporal variations in their toroidal energy, and long-term monitoring will reveal whether this weak magnetism is a transient feature or represents a persistent characteristic.

Overall, our study extends the mass–period diagram (Fig.~\ref{fig:Confusogram}) into previously undersampled regions, providing new insights into the large-scale magnetic properties of M dwarfs. Three rapidly rotating ($P_\mathrm{rot}\sim 1-2$~d) early M dwarfs, CD–29~4446, CD–26~4156, and CD–35~2722, exhibit strong, predominantly poloidal, dipole-dominated fields, consistent with trends observed for slightly less massive stars such as AD~Leo \citep{2008MNRAS.390..567M}. CD–35~2213 represents a peculiar case, with unexpectedly low axisymmetry and a relatively strong toroidal component, atypical for its mass and rotation period, similar to those of AD~Leo. AP~Col shows a dipole-dominated configuration typical of rapidly rotating fully convective stars, but with a weaker and less axisymmetric field, although this is likely partly related to observational coverage. Finally, PM~J05408--3323, a moderately rotating ($P_\mathrm{rot}\sim 17$~d) early M dwarf, stands out for its particularly weak large-scale field, predominantly poloidal and moderately axisymmetric. These observations highlight the diversity of magnetic topologies among M dwarfs and emphasise the need for continued monitoring to assess the stability and temporal evolution of their large-scale fields.

Beyond the mapping of our six M dwarfs in the mass–rotation diagram, our study highlights the importance of monitoring the temporal evolution of their magnetic fields. The diversity of the observed topologies, even among stars with very similar masses and rotation periods, suggests that dynamo regimes may evolve, switch between states, or exhibit cyclic behaviour that cannot be captured through single-epoch observations. Regular spectropolarimetric monitoring is essential to assess the stability of the reconstructed topologies and to reveal potential transitions or cyclic variations in the magnetic field. Future photometric monitoring with PLATO will deliver a more comprehensive view of stellar variability, helping to link magnetic field evolution and long-term photometric activity cycles, and supplying key observational constraints for theoretical models of M‑dwarf magnetism \citep[e.g.][]{2023ApJ...947...36B,2023A&A...678A..82O}.


\begin{acknowledgements}
      We acknowledge financial support from the CNES PLATO grant. Quality observations are made possible by relentless effort of the entire staff at Canada-France-Hawai'i Telescope. We would like to especially thank Luc Arnold, the current CFHT Instrument Scientist for SPIRou. CFHT is located on Maunakea on Hawai'i Island, a mountain of considerable cultural, natural, and ecological significance. Maunakea is a sacred site to Native Hawaiians, also known as Kānaka 'Ōiwi. AAV acknowledges funding from the Dutch Research Council (NWO), with project number VI.C.232.041 of the Talent Programme Vici. SB and AAV acknowledge funding by the Dutch Research Council (NWO) under the project "Exo-space weather and contemporaneous signatures of star-planet interactions" (with project number OCENW.M.22.215 of the research programme "Open Competition Domain Science- M"). CPF acknowledges funding from the European Union's Horizon Europe research and innovation programme under grant agreement No. 101079231 (EXOHOST), and from UK Research and Innovation (UKRI) under the UK government’s Horizon Europe funding guarantee (grant number 10051045). We thank Morgan Deal and Eric Josselin for valuable discussions and helpful comments. 
\end{acknowledgements}

\vspace{1.0cm}
\bibliographystyle{aa}    
\bibliography{references}


\begin{appendix}


\section{Light curves and Lomb-Scargle periodogram for the five others targets}
\label{AppendixA}

\begin{figure}[htbp]
    \centering
    \includegraphics[width=0.80\linewidth]{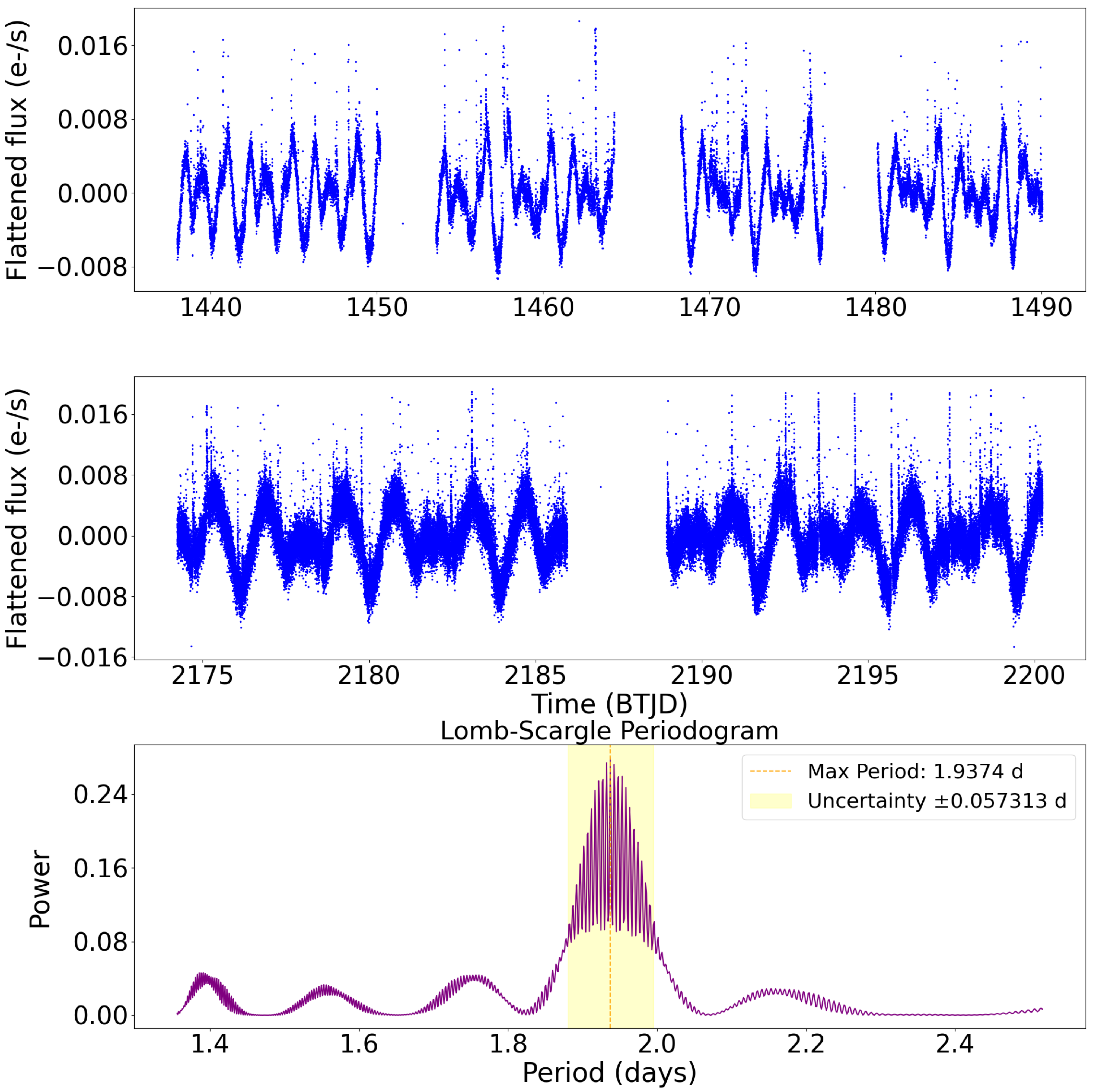}
    \caption{Light curves and Lomb-Scargle periodogram for the star CD-35 2213 observed by \textit{TESS}. The first panel shows the light curve for sectors 0 and 1 combined. The second panel shows the light curve for sector 2. The third panel displays the Lomb-Scargle periodogram computed from all three sectors combined. The light curves exhibit rather frequent flaring, and an estimation of the period uncertainty indicated by the yellow shaded area.}
    \label{fig:CD352213Prot}
\end{figure}

\begin{figure}[htbp]
    \centering
    \includegraphics[width=0.80\linewidth]{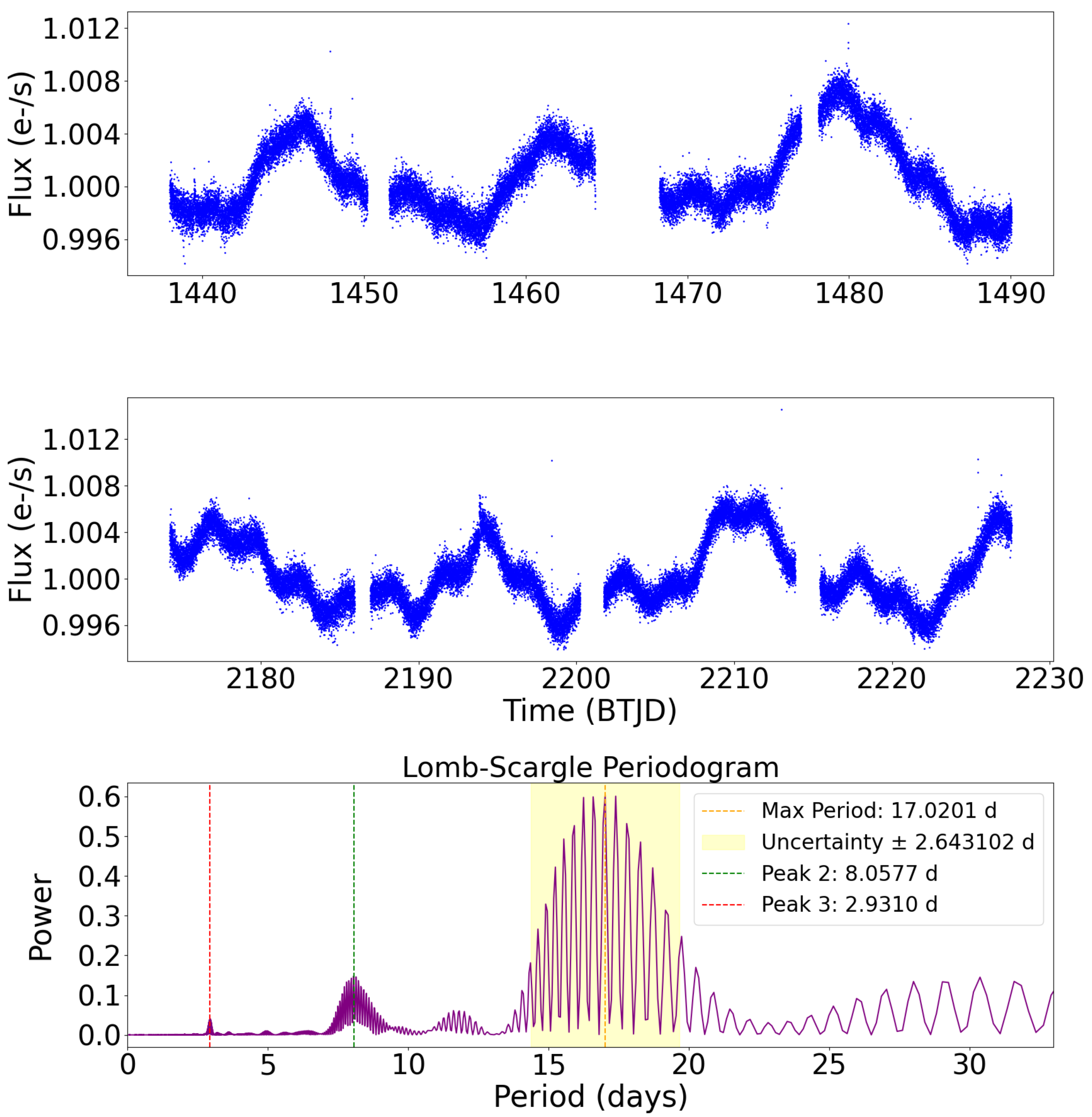}
    \caption{Light curves and Lomb-Scargle periodogram for the star PM J05408-3323 observed by \textit{TESS}. The first panel shows the light curve for sectors 0 and 1 combined. The second panel shows the light curve for sectors 2 and 3 combined. The third panel displays the Lomb-Scargle periodogram computed from all sectors combined, with an estimation of the period uncertainty indicated by the yellow shaded area. The red peak and the green peak highlight additional periodicities discussed in Sect.~5.1.}
    \label{fig:PMJ054083323Prot}
\end{figure}

\begin{figure}[htbp]
    \centering
    \includegraphics[width=\columnwidth]{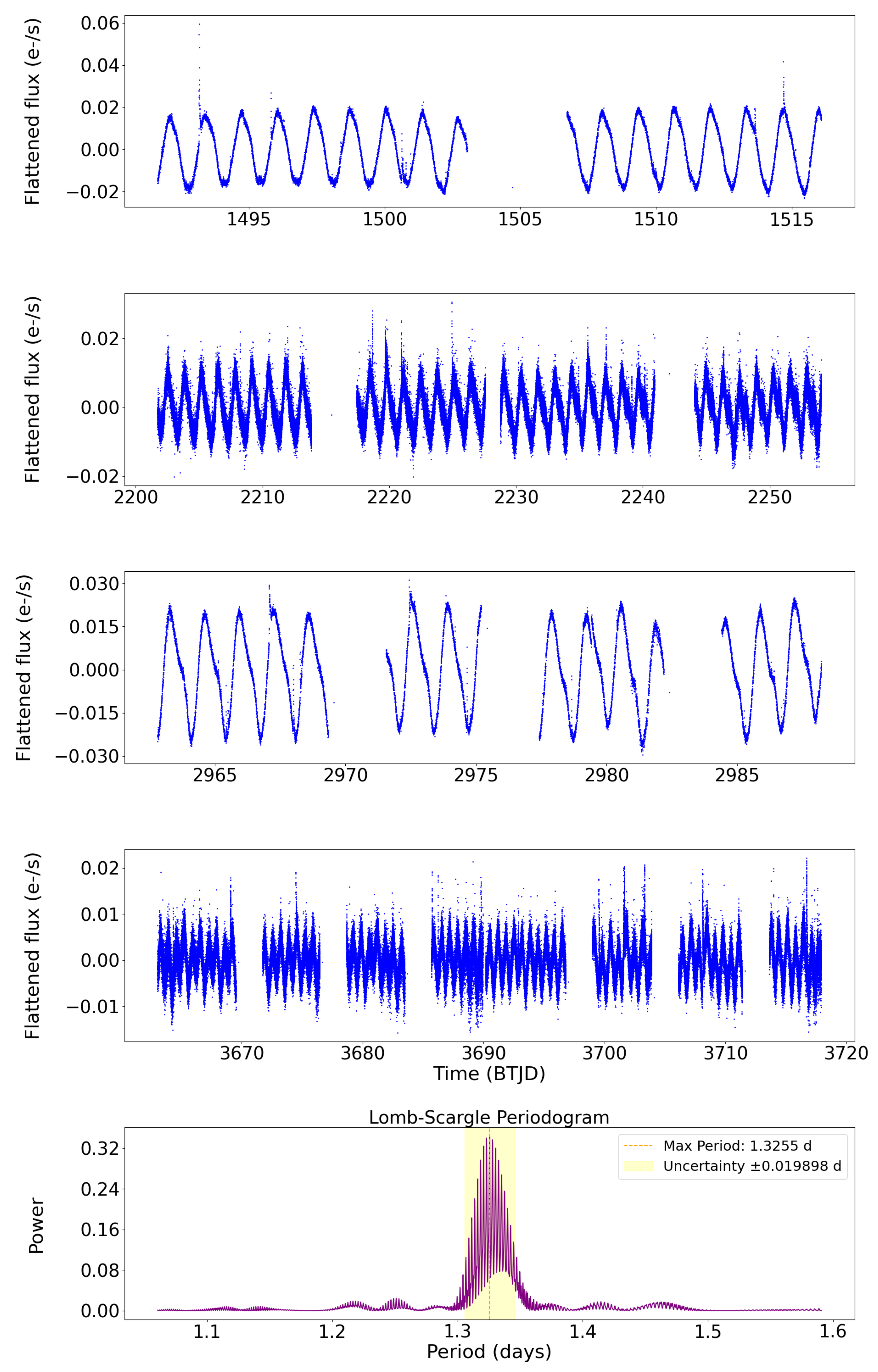}
    \caption{Light curves and Lomb-Scargle periodogram for the star CD-26 4156 observed by \textit{TESS}. The first panel shows the light curve for sector 0. The second panel shows the light curve for sectors 1 and 3 combined. The third panel shows the light curve for sector 5. The fourth panel shows the light curve for sectors 6 and 8 combined. The fifth panel displays the Lomb-Scargle periodogram computed from all sectors combined, with an estimation of the period uncertainty indicated by the yellow shaded area.}
    \label{fig:CD264156Prot}
\end{figure}

\begin{figure}[htbp]
    \centering
    \includegraphics[width=\columnwidth]{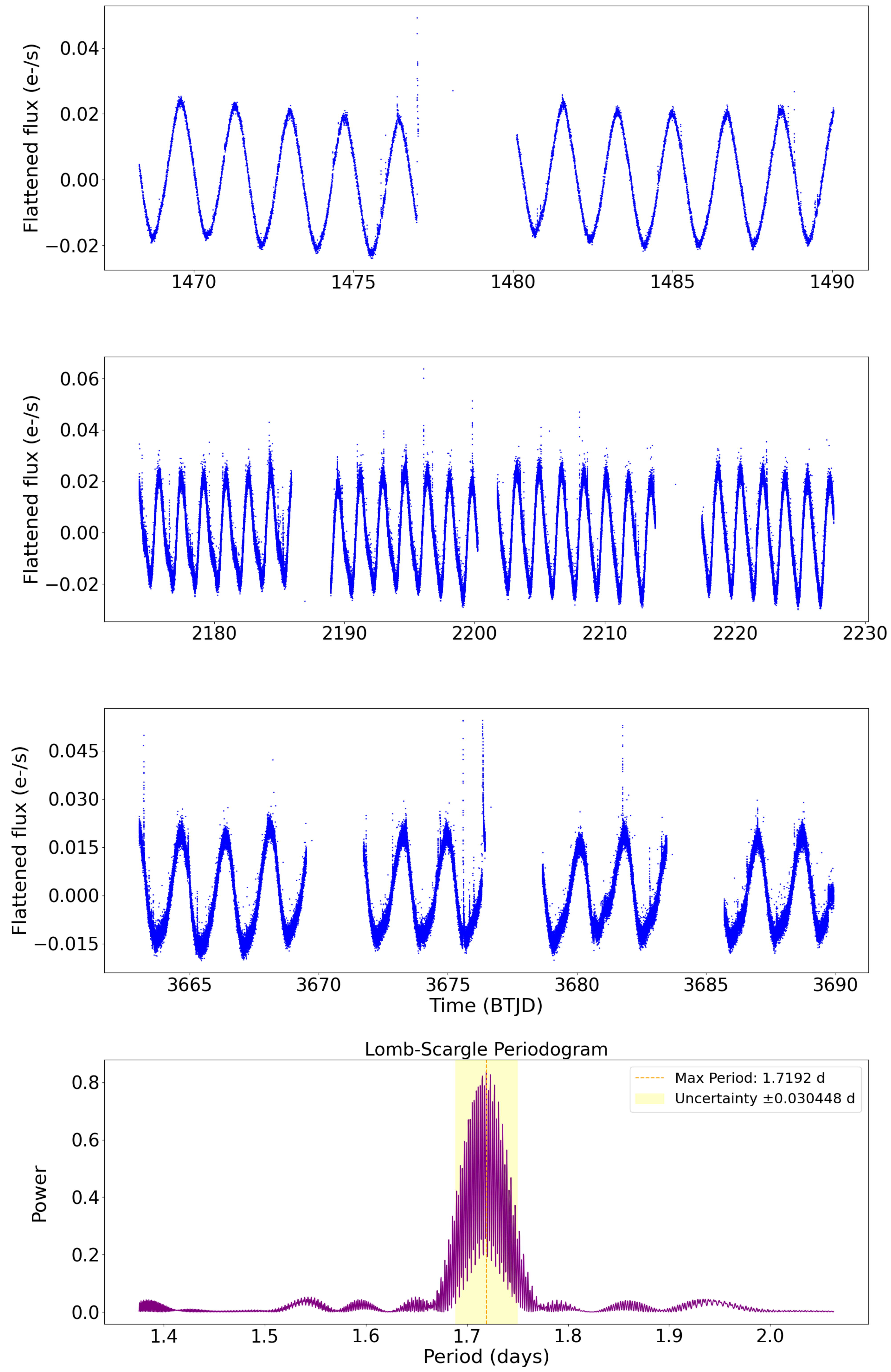}
    \caption{Light curves and Lomb-Scargle periodogram for the star CD-35 2722 observed by \textit{TESS}. The first panel shows the light curve for sector 0. The second panel shows the light curve for sectors 1 and 2 combined. The third panel shows the light curve for sector 5. The fourth panel displays the Lomb-Scargle periodogram computed from all sectors combined, with an estimation of the period uncertainty indicated by the yellow shaded area.}
    \label{fig:CD352722Prot}
\end{figure}

\begin{figure}[htbp]
    \centering
    \includegraphics[width=\columnwidth]{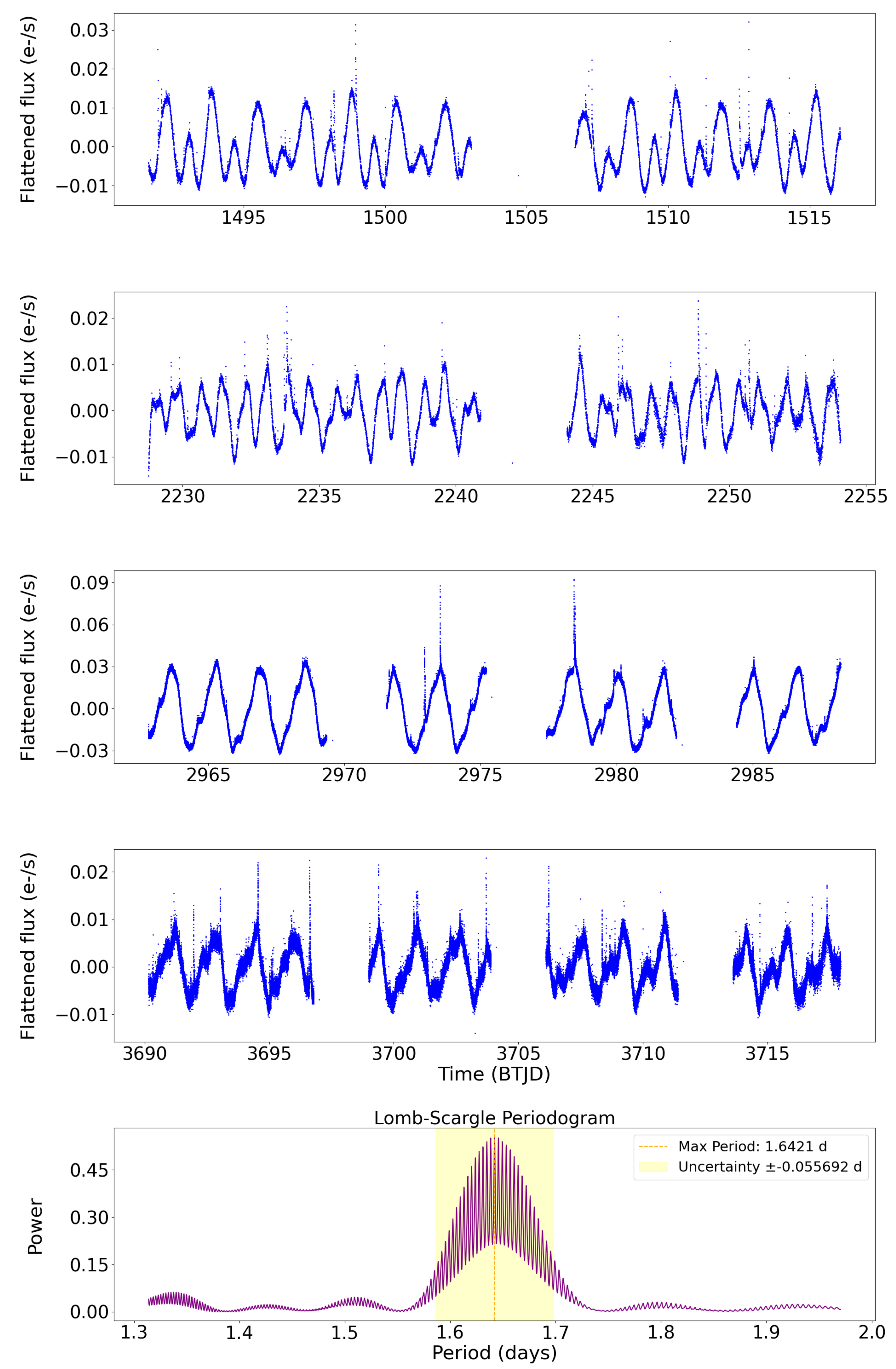}
    \caption{Light curves and Lomb-Scargle periodogram for the star CD-29 4446 observed by \textit{TESS}. The first panel shows the light curve for sector 0. The second panel shows the light curve for sector 1. The third panel shows the light curve for sector 2. The fourth panel shows the light curve for sector 4. The fifth panel displays the Lomb-Scargle periodogram computed from all sectors combined, with an estimation of the period uncertainty indicated by the yellow shaded area.}
    \label{fig:CD294446Prot}
\end{figure}

\clearpage

\onecolumn
\section{}
\label{AppendixB}
\addcontentsline{toc}{section}{} 

\begin{figure}[htbp]
\caption{Fit to the model spectrum of CD-29 4446.}
    \centering
    \includegraphics[width=0.70\linewidth]{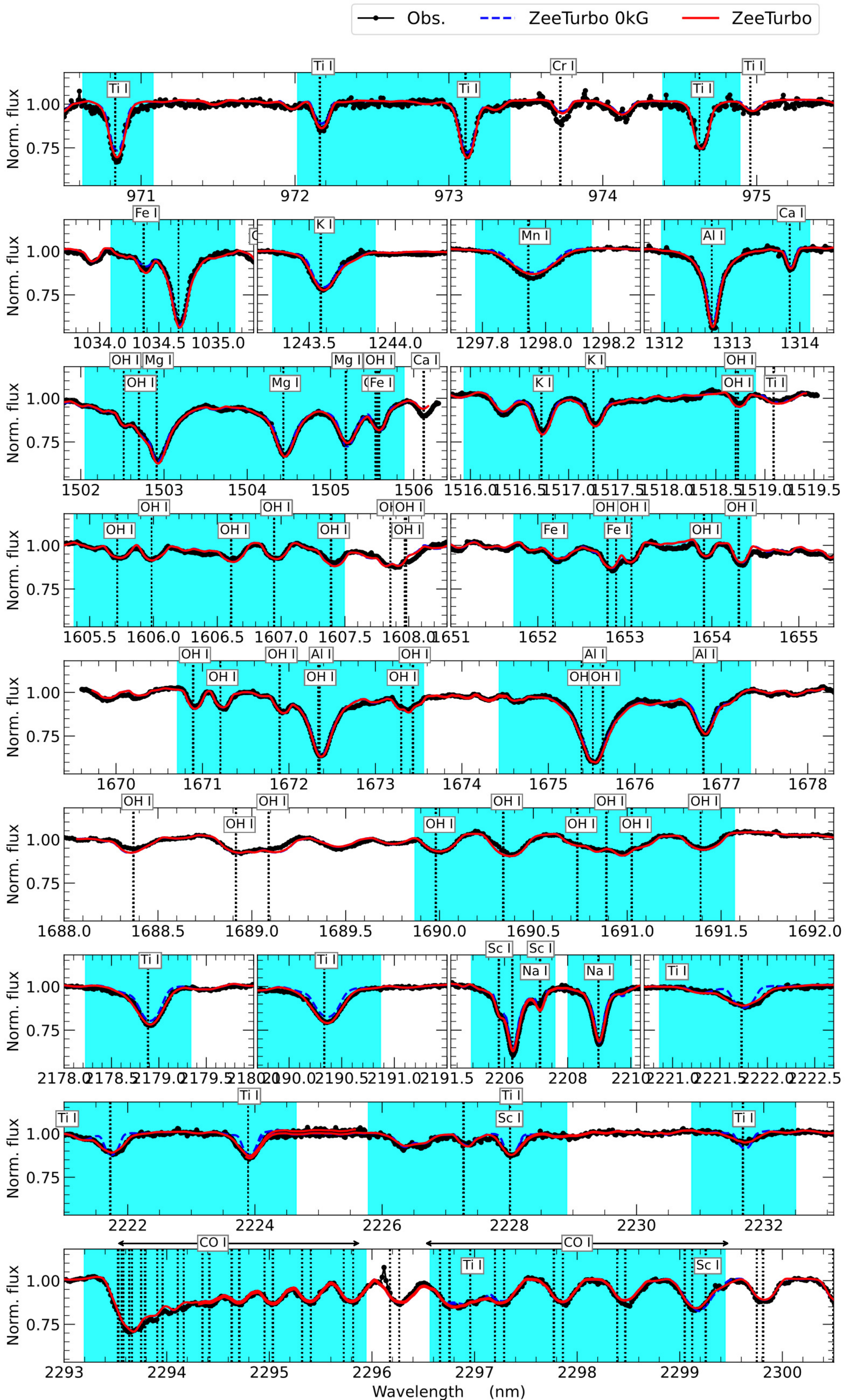}

    \vspace{1mm} 
    {\footnotesize
    \textbf{Notes:} Best-fit model obtained by fitting the observed spectral lines of CD-29 4446 with synthetic spectra computed using ZeeTurbo. 
    The cyan regions indicate the wavelength intervals included in the fitting procedure.
    }
    \label{plotZeeturbo}
\end{figure}

\begin{table*}
\centering
\normalsize
\setlength{\tabcolsep}{4pt}
\caption{Strongest lines used for our fitting procedure.}

\begin{minipage}{0.49\textwidth}
\centering
\begin{tabular}{lcccc}
\hline\hline
Species & $\lambda_{\rm air}$ & $\chi_{\rm low}$ & $\log(gf)$ & $g_{\rm eff}$ [$g_{\rm low}$ -- $g_{\rm up}$] \\
& [$\text{\AA}$] & [eV] & & 

\\
Na I & 22056.422 & 3.191 & -0.072 & 1.17[2.00--1.33] \\
& 22083.684 & 3.191 & -0.518 & 1.33[2.00--0.67] \\
\\
Mg I & 15024.997 & 5.108 & 0.357 & 1.25[2.00--1.50] \\
& 15040.246 & 5.108 & 0.135 & 1.75[2.00--1.50] \\
& 15047.714 & 5.108 & -0.341 & 2.00[2.00--0.00] \\
\\
Al I & 13123.434 & 3.143 & -0.156 & 1.17[2.00--1.33] \\
& 16718.963 & 4.085 & -0.638 & 0.83[0.67--0.80] \\
& 16750.551 & 4.087 & -0.796 & 1.10[1.33--1.20] \\
& 16763.369 & 4.087 & -1.578 & 1.07[1.33--0.80] \\
\\
K I & 12432.277 & 1.610 & -0.944 & 1.33[0.67--2.00] \\
& 15163.067 & 2.670 & 0.689 & 1.06[1.20--1.14] \\
& 15168.376 & 2.670 & 0.480 & 0.90[0.80--0.86] \\
\\
Ca I & 10343.819 & 2.932 & -0.300 & 1.00[1.00--0.00] \\
& 13134.941 & 4.451 & 0.085 & 1.25[1.25--1.25] \\
\\
Sc I & 22052.109 & 1.448 & -1.616 & 1.16[1.33--1.43] \\
& 22065.277 & 1.439 & -2.596 & 1.11[1.24--1.34] \\
& 22266.740 & 1.428 & -1.872 & 0.50[0.40--0.00] \\
& 22986.236 & 1.448 & -2.737 & 1.33[1.33--1.33] \\
\\
Ti I & 9705.665 & 0.826 & -1.009 & 1.25[1.25--1.26] \\
& 9718.960 & 1.503 & -1.181 & 0.95[0.98--1.00] \\
& 9728.405 & 0.818 & -1.206 & 0.99[0.99--1.00] \\
& 9743.605 & 0.813 & -1.306 & 0.00[0.00--0.00] \\
& 21782.920 & 1.749 & -1.170 & 1.29[1.66--1.51] \\
& 21897.377 & 1.739 & -1.470 & 1.16[1.82--1.49] \\
& 22211.219 & 1.734 & -1.780 & 2.08[2.50--1.65] \\
& 22232.844 & 1.739 & -1.690 & 1.66[1.82--1.50] \\
& 22274.006 & 1.749 & -1.800 & 1.57[1.66--1.49] \\
& 22310.611 & 1.734 & -2.071 & 2.50[2.50--0.00] \\
& 22963.330 & 1.887 & -1.530 & 1.11[1.21--1.26] \\
\\
Mn I & 12975.929 & 2.888 & -1.797 & 1.22[1.43--1.60] \\
\\
Fe I & 10340.885 & 2.198 & -3.577 & 0.68[1.82--1.25] \\
& 15051.749 & 5.352 & 0.426 & 1.52[1.66--1.75] \\
& 16517.223 & 6.286 & 0.679 & 1.18[1.36--1.31] \\
& 16524.467 & 6.336 & 0.688 & 0.96[1.38--1.26] \\
\\

OH & 15021.039 & 0.127 & -5.595 & ... \\
& 15022.865 & 0.127 & -5.595 & ... \\
& 15051.327 & 0.442 & -5.753 & ... \\
& 15051.541 & 0.442 & -5.753 & ... \\
& 15182.857 & 0.465 & -5.969 & ... \\
& 15183.123 & 0.465 & -5.969 & ... \\
& 16052.766 & 0.639 & -4.976 & ... \\
& 16055.466 & 0.639 & -4.976 & ... \\
& 16061.702 & 0.476 & -5.222 & ... \\

\hline
\end{tabular}
\end{minipage}
\hfill
\begin{minipage}{0.49\textwidth}
\centering
\begin{tabular}{lcccc}
\hline\hline
Species & $\lambda_{\rm air}$ & $\chi_{\rm low}$ & $\log(gf)$ & $g_{\rm eff}$ [$g_{\rm low}$ -- $g_{\rm up}$] \\
& [$\text{\AA}$] & [eV] & & 

\\

OH & 16065.053 & 0.477 & -5.222 & ... \\
& 16069.525 & 0.472 & -5.191 & ... \\
& 16523.498 & 0.786 & -4.842 & ... \\
& 16526.254 & 0.786 & -4.842 & ... \\
& 16534.582 & 0.781 & -4.806 & ... \\
& 16538.588 & 0.782 & -4.806 & .... \\
& 16704.359 & 0.841 & -4.791 & .... \\
& 16714.359 & 0.837 & -4.758 & ... \\
& 16718.855 & 0.838 & -4.758 & ... \\
& 16749.255 & 1.015 & -4.741 & ... \\
& 16751.725 & 1.016 & -4.741 & ... \\
& 16895.183 & 0.900 & -4.743 & ... \\
& 16898.778 & 0.901 & -4.743 & ... \\
& 16902.733 & 1.053 & -4.674 & ... \\
& 16904.278 & 0.896 & -4.712 & ... \\
& 16909.289 & 0.897 & -4.712 & .... \\
\\
CO I & 22928.98949 & 0.627 & -5.202 & ... \\
& 22929.04582 & 0.603 & -5.214 & ... \\
& 22929.34121 & 0.651 & -5.191 & ... \\
& 22929.50944 & 0.579 & -5.225 & ... \\
& 22930.10180 & 0.676 & -5.180 & ... \\
& 22930.37964 & 0.556 & -5.237 & ... \\
& 22931.27209 & 0.701 & -5.169 & ... \\
& 22931.65575 & 0.534 & -5.249 & ... \\
& 22932.85299 & 0.727 & -5.158 & ... \\
& 22933.33713 & 0.512 & -5.261 & ... \\
& 22934.84543 & 0.753 & -5.147 & ... \\
& 22935.42321 & 0.490 & -5.274 & ... \\
& 22937.25040 & 0.780 & -5.137 & ... \\
& 22937.91342 & 0.469 & -5.286 & ... \\
& 22940.06893 & 0.807 & -5.126 & ... \\
& 22940.80729 & 0.448 & -5.299 & ... \\
& 22943.30208 & 0.834 & -5.116 & ... \\
& 22944.10433 & 0.428 & -5.312 & ... \\
& 22946.95100 & 0.862 & -5.105 & ... \\
& 22947.80414 & 0.408 & -5.325 & ... \\
& 22951.01685 & 0.891 & -5.095 & ... \\
& 22951.90634 & 0.389 & -5.339 & ... \\
& 22955.50085 & 0.920 & -5.085 & ... \\
& 22956.41060 & 0.370 & -5.352 & ... \\
& 22960.40427 & 0.949 & -5.075 & ... \\
& 22961.31662 & 0.352 & -5.366 & ... \\
& 22965.72841 & 0.979 & -5.066 & ... \\
& 22966.62415 & 0.334 & -5.381 & ... \\
& 22971.47466 & 1.009 & -5.056 & ... \\
& 22972.33298 & 0.316 & -5.395 & ... \\
& 22977.64442 & 1.039 & -5.046 & ... \\
& 22978.44294 & 0.299 & -5.410 & ... \\
& 22984.23916 & 1.070 & -5.037 & ... \\
& 22984.95390 & 0.283 & -5.425 & ... \\
\hline

\label{lines}
\end{tabular}
\end{minipage}

\vspace{6mm}
{\footnotesize
\textbf{Notes:} $\lambda_{\rm air}$ represents the wavelength in air, $\chi_{\rm low}$ is the excitation potential, $\log (gf)$ is the oscillator strength, and $g_{\rm eff}$, $g_{\rm low}$, $g_{\rm up}$ are the effective, lower, and upper Landé factors. Landé factors are not displayed for molecular lines as ZeeTurbo does not currently handle molecular Zeeman splitting.}
\end{table*}

\clearpage

\twocolumn
\section{Longitudinal magnetic field measurements}
\label{AppendixC}

\begin{figure}[htbp]
    \centering
    \includegraphics[width=\columnwidth]{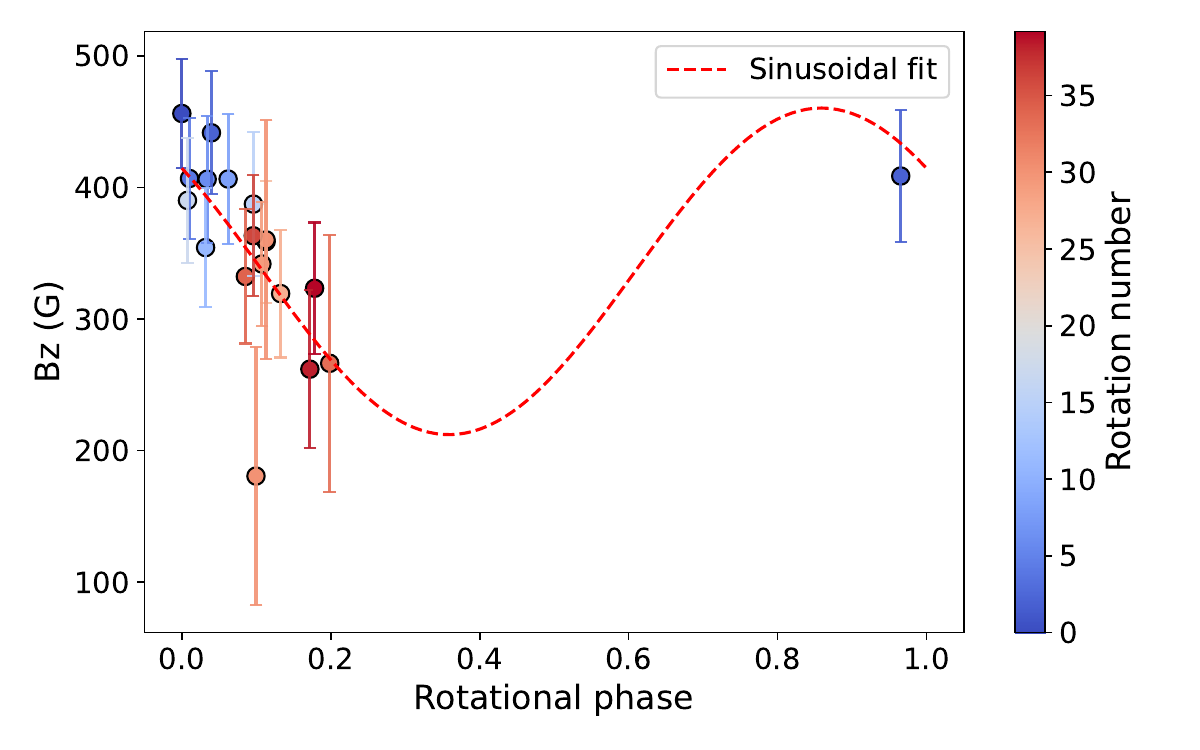}
    \caption*{(a) AP Col}
\end{figure}

\begin{figure}[htbp]
    \centering
    \includegraphics[width=\columnwidth]{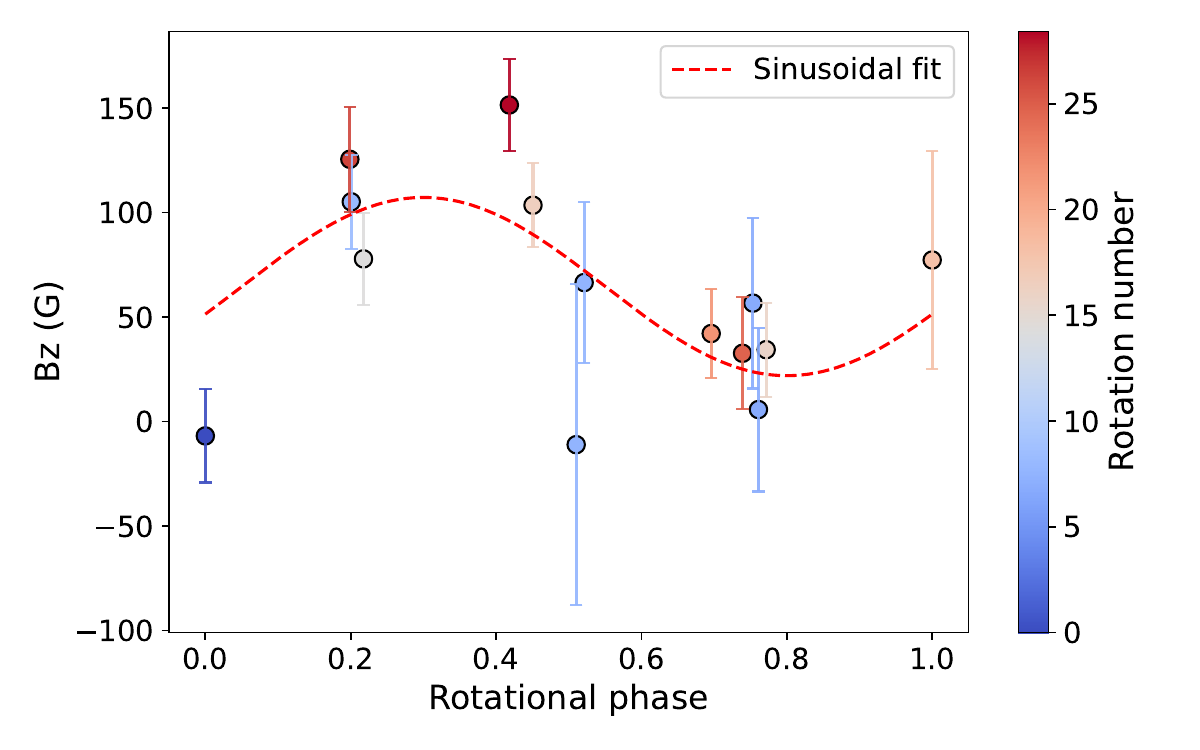}
    \caption*{(b) CD-26 4156}
\end{figure}

\begin{figure}[htbp]
    \centering
    \includegraphics[width=\columnwidth]{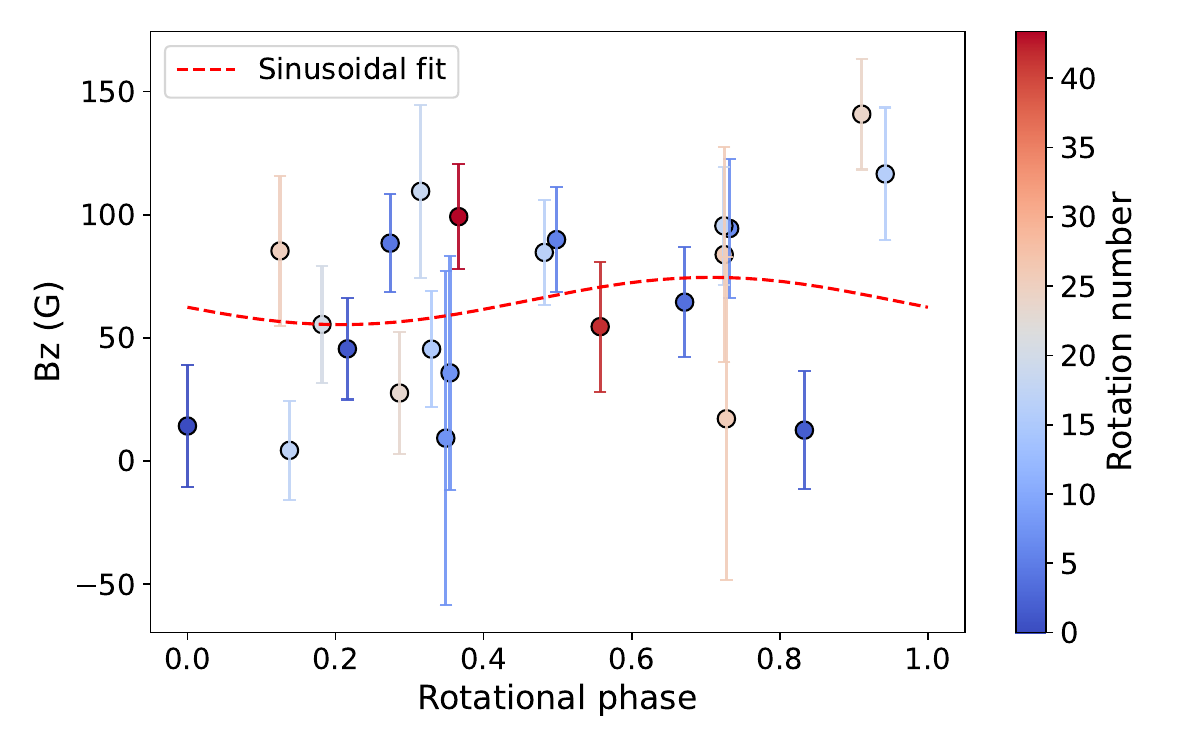}
    \caption*{(c) CD-29 4446}
\end{figure}

\begin{figure}[htbp]
    \centering
    \includegraphics[width=\columnwidth]{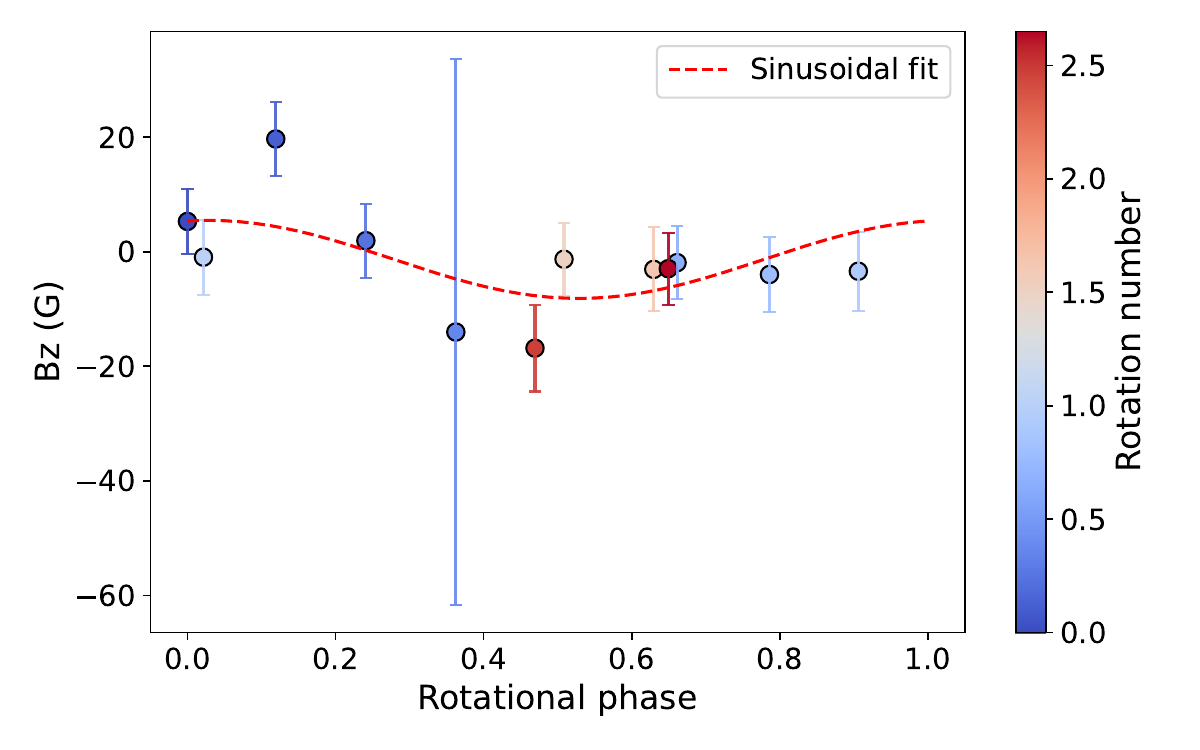}
    \caption*{(d) PM J05408-3323}
    \caption*{Fig. C: Longitudinal magnetic field $B_{z}$ as a function of rotational phase for the targets shown in panels (a)--(d). The colour bar indicates the number of rotations.}
\end{figure}

\clearpage

\twocolumn
\section{Stokes V profiles of all targets}
\label{AppendixD}

\begin{figure}[htbp]
\centering
\includegraphics[width=\columnwidth]{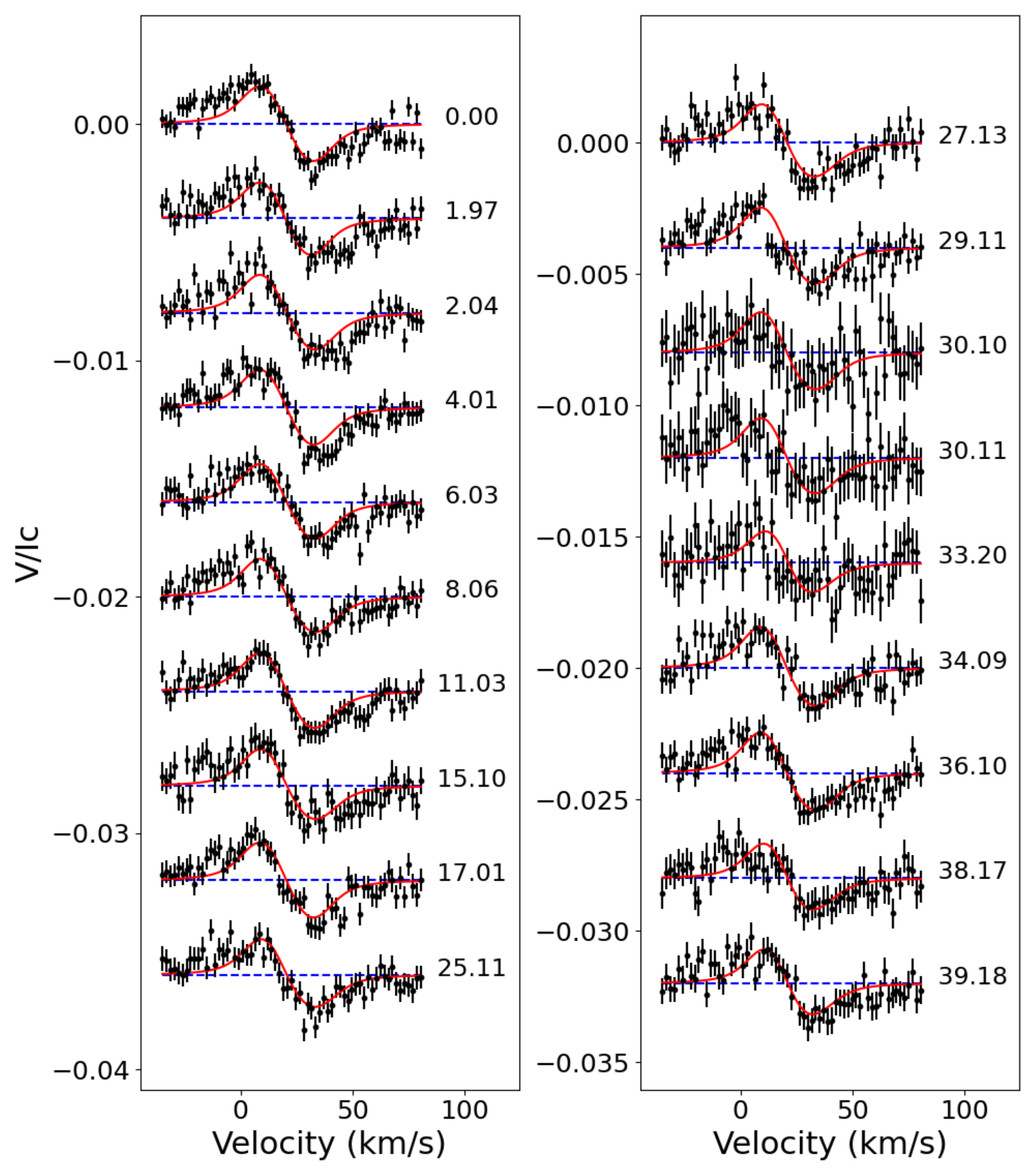}
\caption*{(a) AP Col}
\label{fig:APCol_StokesV}
\end{figure}

\begin{figure}[htbp]
\centering
\includegraphics[width=\columnwidth]{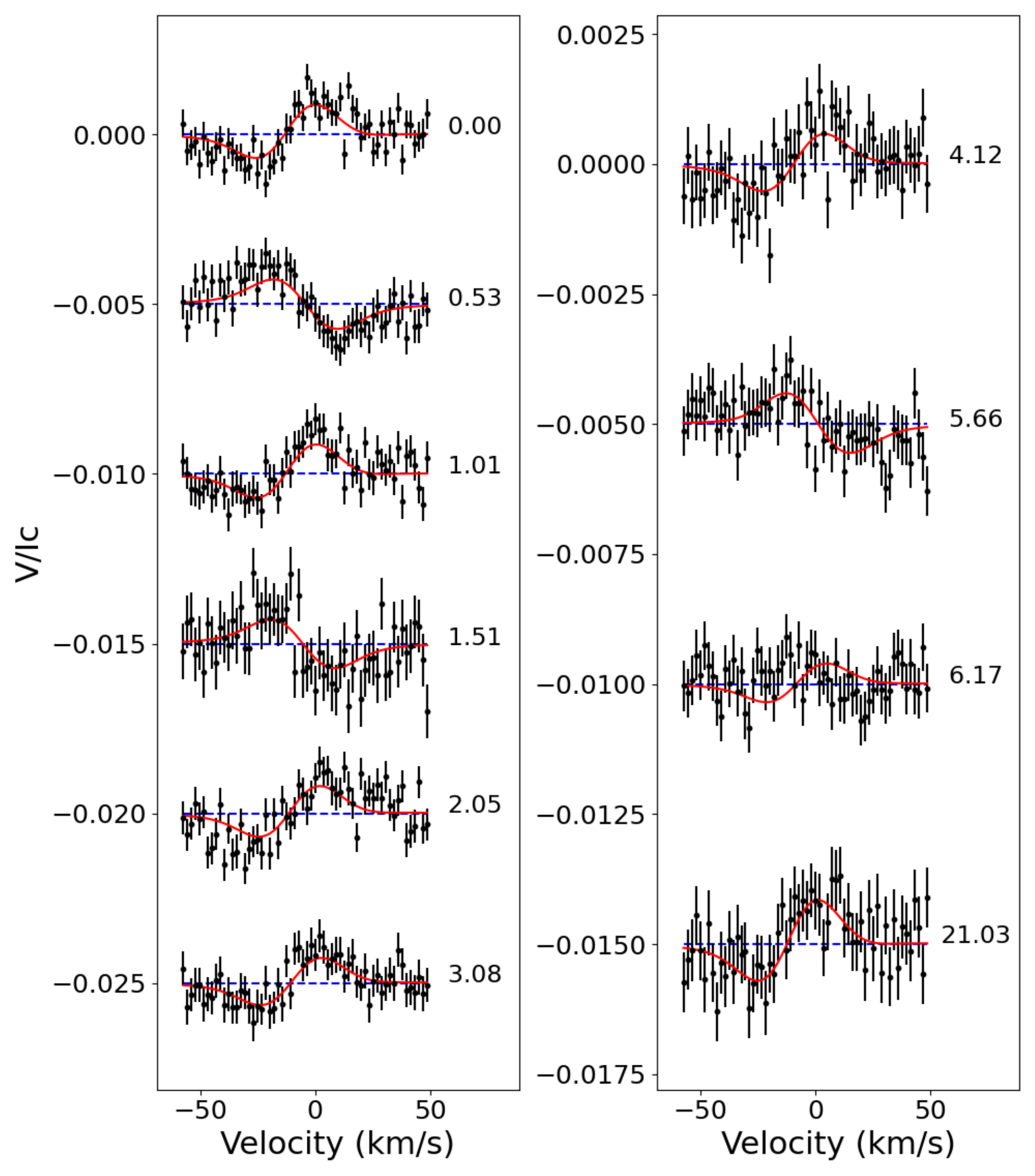}
\caption*{(b) CD-35 2213}
\label{fig:CD352213_StokesV}
\end{figure}

\begin{figure}[htbp]
\centering
\includegraphics[width=\columnwidth]{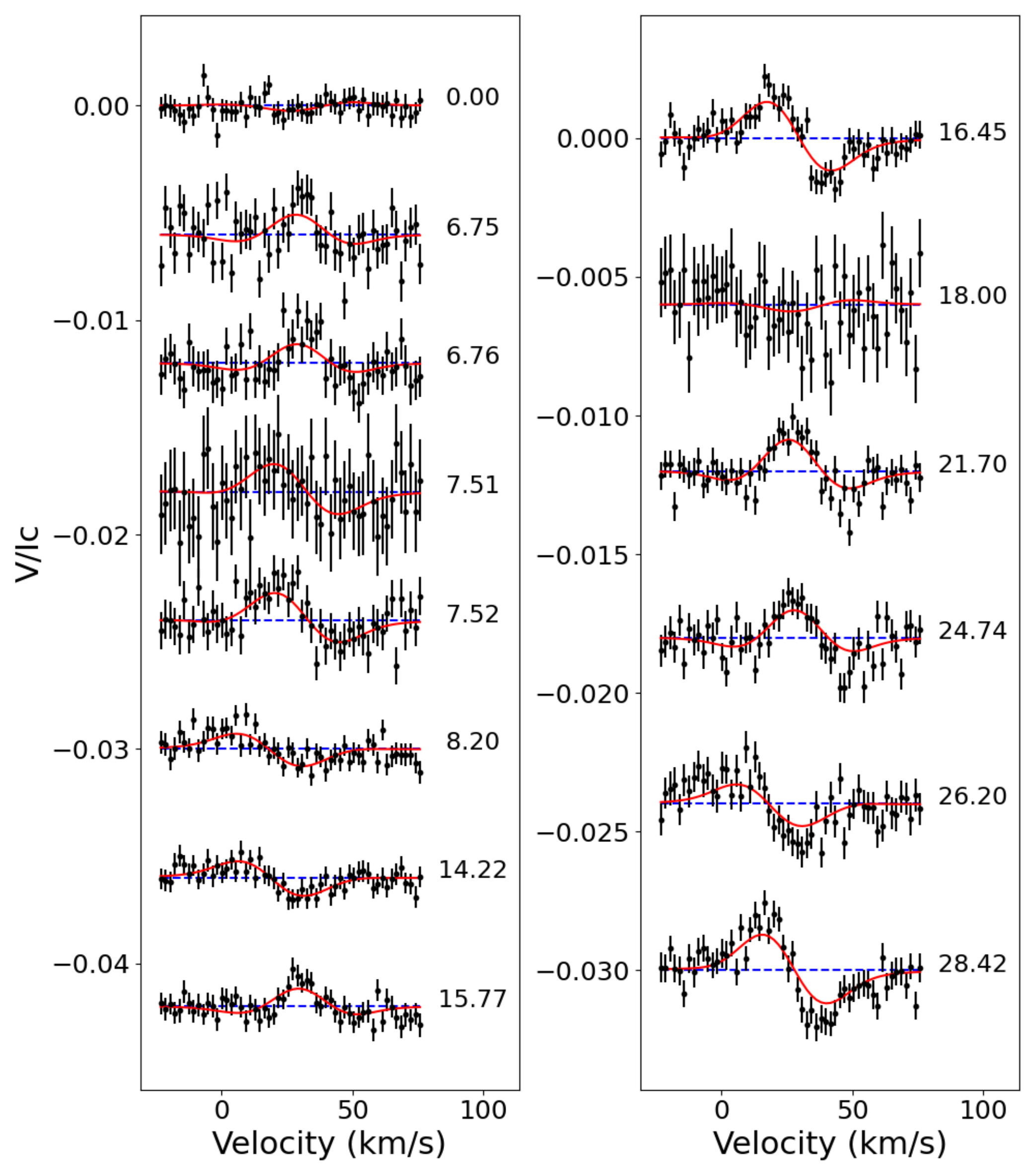}
\caption*{(c) CD-26 4156}
\label{fig:CD264156_StokesV}
\end{figure}

\begin{figure}[htbp]
\centering
\includegraphics[width=\columnwidth]{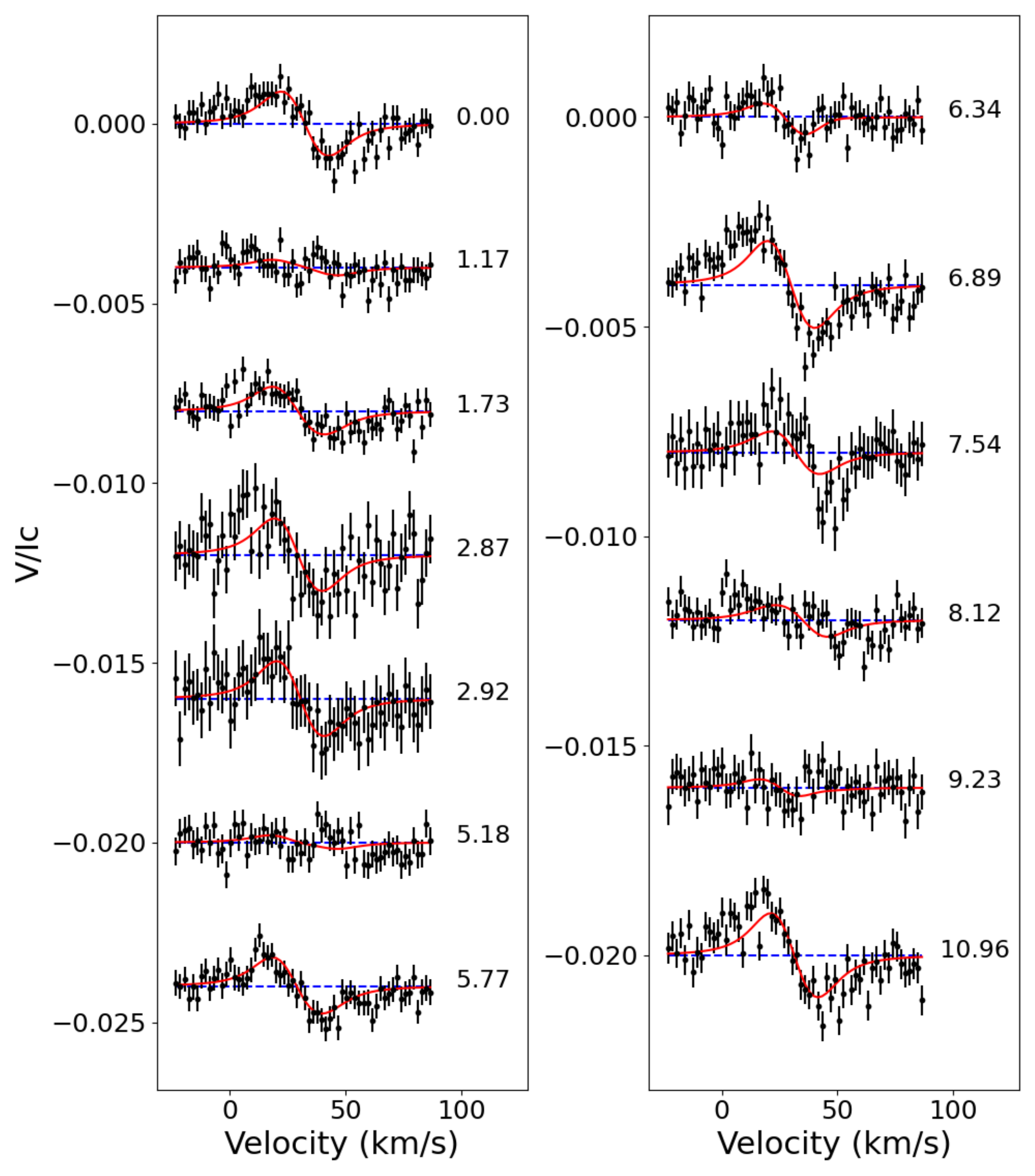}
\caption*{(d) CD-35 2722}
\label{fig:CD352722_StokesV}
\end{figure}

\begin{figure}[htbp]
\centering
\includegraphics[width=\columnwidth]{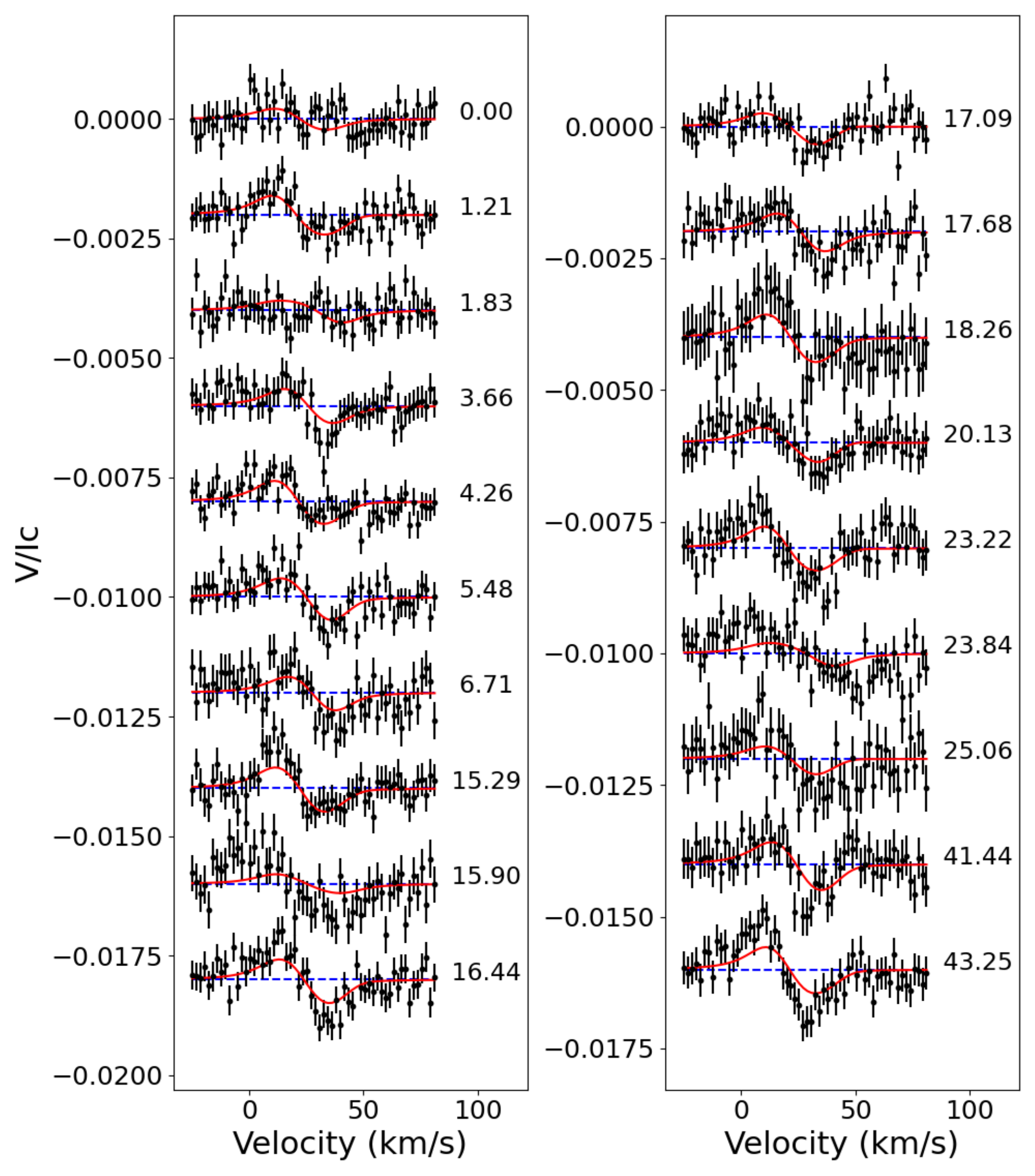}
\caption*{(e) CD-29 4446}
\label{fig:CD294446_StokesV}
\end{figure}

\begin{figure}[htbp]
\centering
\includegraphics[width=\columnwidth]{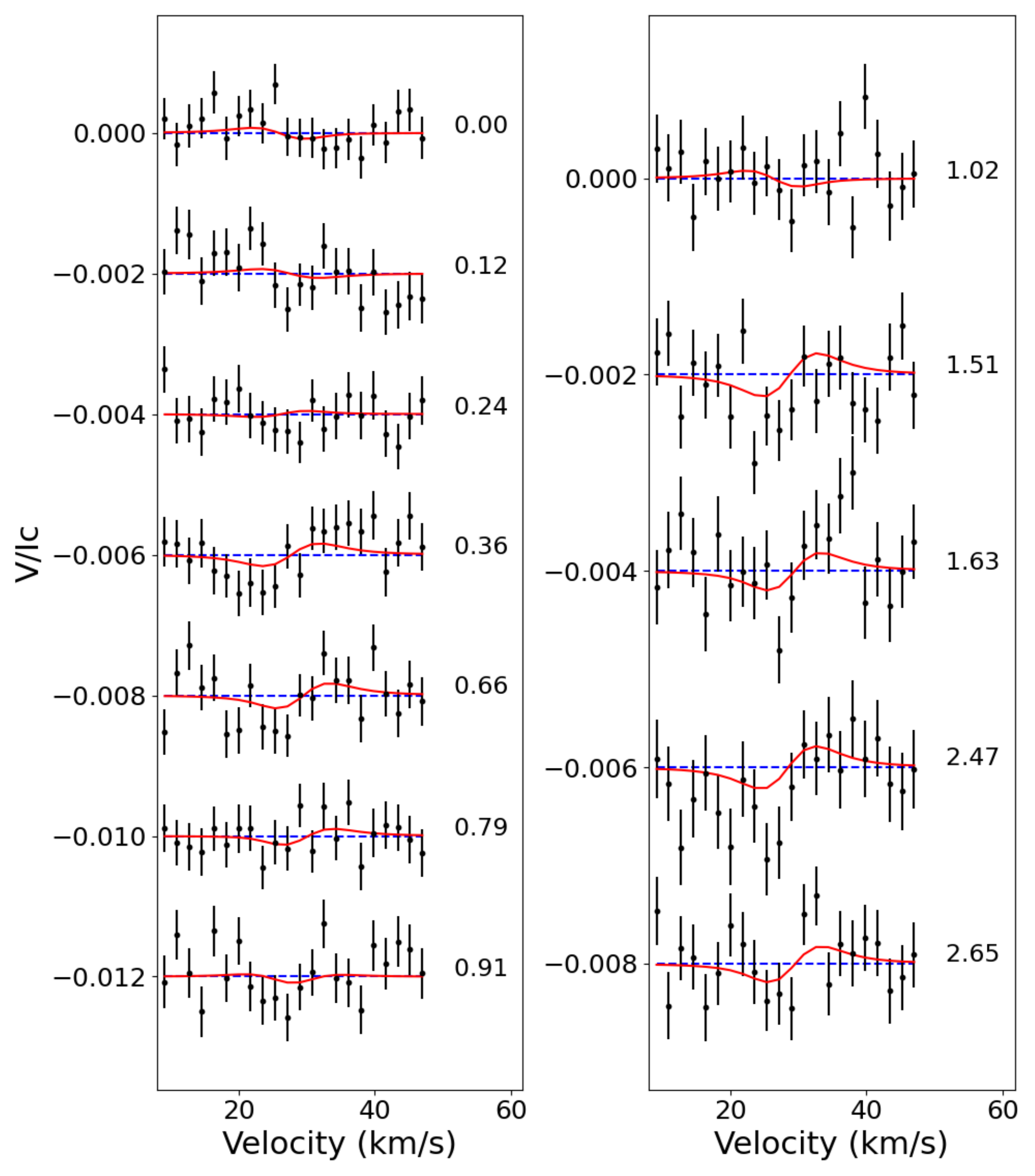}
\caption*{(f) PM J05408-3323}
\caption*{Stokes $V$ profiles of all targets obtained with the ZDI code. 
The observed profiles are shown in black, while the red lines correspond to the best-fit models computed by ZDI. This convention applies to all six subpanels (a)--(f).}
\label{fig:PMJ054083323_StokesV}
\end{figure}

\clearpage

\onecolumn
\section{ Observing log (2024)}
\label{AppendixE}

\begin{table}[htbp]
\centering
\caption{Observing log of the six M dwarfs with SPIRou. The columns are (1) date, (2) universal time, (3) heliocentric Julian date, (4) rotation cycle computed from the stellar rotation period, (5) exposure time per polarimetric sequence, (6) signal-to-noise ratio per sequence, and (7) longitudinal magnetic field values for each observation, with associated error bars.}
\begin{tabular}{lccccccc}
\hline
Date (2024) & UT [hh:mm:ss] & HJD [$-$2450000] & $n_{\mathrm{cyc}}$ & $t_{\mathrm{exp}}$ [s] & S/N & $B_{\mathrm{z}}$ [G] \\
\hline
\multicolumn{7}{c}{\textbf{AP Col}\rule{0pt}{2.5ex}} \\
\hline
November 09 & 12:41:01.34 & 60623.5284877 & 0.00 & 4x184 & 119 & $456.2 \pm 42$\\
November 11 & 11:34:16.49 & 60625.4821353 & 1.97 & 4x163 & 94 & $408.7 \pm 50$\\
November 11 & 13:20:39.52 & 60625.5560129 & 2.04 & 4x167 & 101 & $441.5 \pm 47$\\
November 13 & 12:21:13.05 & 60627.5147344 & 4.01 & 4x146 & 100 & $406.8 \pm 46$\\
November 15 & 12:37:53.65 & 60629.5263154 & 6.03 & 4x174 & 97 & $406.2 \pm 48$\\
November 17 & 13:00:48.23 & 60631.5422249 & 8.06 & 4x176 & 102 & $406.3 \pm 49$\\
November 20 & 11:51:38.20 & 60634.4941922 & 11.03 & 4x159 & 102 & $354.3 \pm 45$\\
November 24 & 12:48:59.81 & 60638.5340256 & 15.10 & 4x182 & 68 & $387.3 \pm 55$\\
November 26 & 10:24:56.33 & 60640.4339853 & 17.01 & 4x162 & 102 & $390.1 \pm 48$\\
December 04 & 11:47:01.52 & 60648.4909898 & 25.11 & 4x135 & 103 & $358.8 \pm 46$\\
December 06 & 11:57:38.41 & 60650.4983613 & 27.13 & 4x182 & 101 & $319.3 \pm 49$\\
December 08 & 11:04:23.55 & 60652.4613837 & 29.11 & 4x166 & 100 & $341.8 \pm 47$\\
December 09 & 10:44:20.26 & 60653.4474568 & 30.01 & 4x184 & 48 & $180.6 \pm 98$\\
December 09 & 11:03:40.60 & 60653.4608866 & 30.11 & 4x184 & 49 & $360.1 \pm 91$\\
December 12 & 12:40:10.73 & 60656.5279020 & 33.20 & 4x184 & 54 & $266.3 \pm 98$\\
December 13 & 09:49:23.25 & 60657.4092969 & 34.09 & 4x171 & 101 & $332.3 \pm 51$\\
December 15 & 09:47:13.39 & 60659.4077938 & 36.10 & 4x138 & 101 & $363.2 \pm 46$\\
December 17 & 11:18:48.34 & 60661.4713928 & 38.17 & 4x184 & 69 & $261.9 \pm 60$\\
December 18 & 11:19:13.23 & 60662.4716809 & 39.18 & 4x163 & 100 & $323.3 \pm 50$\\
\hline
\multicolumn{7}{c}{ \textbf{CD-35 2213}\rule{0pt}{2.5ex}} \\
\hline
November 09 & 12:13:48.36 & 60623.5095875 & 0.00 & 4x128 & 114 & $-167.4 \pm 31$ \\
November 10 & 13:05:37.93 & 60624.5455778 & 0.52 & 4x97 & 100 & $221.8 \pm 36$\\
November 11 & 11:19:33.66 & 60625.4719173 & 0.98 & 4x117 & 103 & $-143.3 \pm 36$ \\
November 12 & 10:35:18.98 & 60626.4411919 & 1.47 & 4x128 & 63 & $140.3 \pm 59$\\
November 13 & 11:56:20.95 & 60627.4974647 & 2.00 & 4x104 & 101 & $-197.5 \pm 35$ \\
November 15 & 11:40:54.18 & 60629.4867381 & 3.00 & 4x125 & 94 & $-115.1 \pm 37$\\
November 17 & 12:30:28.22 & 60631.5211599 & 4.02 & 4x128 & 89 & $-146.5 \pm 40$\\
November 20 & 12:20:30.02 & 60634.5142364 & 5.52 & 4x123 & 103 & $161.9 \pm 35$\\
November 21 & 12:06:55.88 & 60635.5048135 & 6.02 & 4x110 & 100 & $17.3 \pm 34$\\
November 22 & 12:26:00.53 & 60636.5180617 & 6.52 & 4x117 & 103 & $6.6 \pm 34$\\
December 17 & 11:06:41.69 & 60661.4629826 & 19.03 & 4x128 & 80 & $119.7 \pm 49$\\
December 20 & 08:55:58.26 & 60664.3722021 & 20.49 & 4x127 & 101 & $-120.2 \pm 43$ \\
December 23 & 08:27:47.56 & 60667.3526338 & 21.99 & 4x100 & 101 & $107.7 \pm 36$\\
\hline
\multicolumn{7}{c}{ \textbf{CD-26 4156}\rule{0pt}{2.5ex}} \\
\hline
November 15 & 14:55:17.49 & 60629.6217302 & 0.00 & 4x184 & 90 & $-6.8 \pm 22$\\
November 24 & 14:16:34.44 & 60638.5948430 & 6.75 & 4x184 & 44 & $56.7 \pm 41$ \\
November 24 & 14:31:07.50 & 60638.6049480 & 6.76 & 4x184 & 72 & $5.8 \pm 39$\\
November 25 & 14:24:43.40 & 60639.6005023 & 7.51 & 4x184 & 38 & $-11.0 \pm 77$\\
November 25 & 14:45:42.33 & 60639.6150732 & 7.52 & 4x184 & 57 & $66.5 \pm 39$\\
November 26 & 12:26:01.07 & 60640.5180680 & 8.20 & 4x184 & 95 & $105.2 \pm 22$ \\
December 04 & 12:18:11.95 & 60648.5126383 & 14.22 & 4x169 & 101 & $77.9 \pm 22$\\
December 06 & 13:51:26.28 & 60650.5773875 & 15.77 & 4x184 & 92 & $34.4 \pm 23$\\
December 07 & 11:30:59.97 & 60651.4798608 & 16.45 & 4x176 & 102 & $103.5 \pm 20$\\
December 09 & 12:55:06.64 & 60653.5382713 & 18.00 & 4x184 & 43 & $77.3 \pm 52$\\
December 14 & 10:46:42.92 & 60658.4491079 & 21.70 & 4x184 & 101 & $42.2 \pm 21$ \\
December 18 & 11:49:10.60 & 60662.4924838 & 24.74 & 4x184 & 87 & $32.7 \pm 27$\\
December 20 & 10:22:11.34 & 60664.4320757 & 26.20 & 4x178 & 87 & $125.5 \pm 25$\\
December 23 & 09:08:41.04 & 60667.3810305 & 28.42 & 4x181 & 94 & $151.5 \pm 22$\\
\hline
\end{tabular}
\end{table}

\begin{table}[htbp]
\centering
\begin{tabular}{lccccccc}
\hline
Date (2024) & UT [hh:mm:ss] & HJD [$-$2450000] & $n_{\mathrm{cyc}}$ & $t_{\mathrm{exp}}$ [s] & S/N & $B_{\mathrm{z}}$ [G]\\
\hline
\multicolumn{7}{c}{ \textbf{CD-35 2722}\rule{0pt}{2.5ex}} \\
\hline
December 04 & 11:59:07.08 & 60648.4993875 & 0.00 & 4x144 & 102 & $173.0 \pm 25$\\
December 06 & 12:12:40.86 & 60650.5088063 & 1.17 & 4x182 & 91 & $81.9 \pm 25$\\
December 07 & 11:17:46.34 & 60651.4706752 & 1.73 & 4x148 & 102 & $127.3 \pm 24$ \\
December 09 & 10:28:25.45 & 60653.4364057 & 2.87 & 4x184 & 63 & $145.6 \pm 48$\\
December 09 & 12:39:20.01 & 60653.5273150 & 2.92 & 4x184 & 55 & $173.9 \pm 53$\\
December 13 & 10:04:03.95 & 60657.4194902 & 5.18 & 4x184 & 85 & $62.7 \pm 27$\\
December 14 & 10:33:55.40 & 60658.4402245 & 5.77 & 4x148 & 100 & $127.2 \pm 25$\\
December 15 & 9:59:51.24 & 60659.4165653 & 6.34 & 4x152 & 102 & $59.1 \pm 24$\\
December 16 & 8:48:07.34 & 60660.3667516 & 6.89 & 4x177 & 101 & $236.1 \pm 25$\\
December 17 & 11:34:10.50 & 60661.4820660 & 7.54 & 4x184 & 86 & $89.7 \pm 37$\\
December 18 & 11:33:19.45 & 60662.4814751 & 8.12 & 4x182 & 95 & $99.3 \pm 26$\\
December 20 & 9:24:01.67 & 60664.3916860 & 9.23 & 4x184 & 82 & $59.1 \pm 30$\\
December 23 & 8:51:08.13 & 60667.3688441 & 10.96 & 4x164 & 101 & $212.8 \pm 26$\\
\hline
\multicolumn{7}{c}{ \textbf{CD-29 4446}\rule{0pt}{2.5ex}} \\
\hline
October 13 & 15:41:13.84 & 60596.6536324 & 0.00 & 4x61 & 106 & $14.2 \pm 25$ \\
October 15 & 15:17:34.50 & 60598.6372048 & 1.21 & 4x61 & 115 & $45.5 \pm 21$ \\
October 16 & 15:27:24.12 & 60599.6440291 & 1.83 & 4x61 & 109 & $12.5 \pm 24$\\
October 19 & 15:25:18.18 & 60602.6425715 & 3.66 & 4x61 & 110 & $64.5 \pm 22$ \\
October 20 & 15:00:00.41 & 60603.6250048 & 4.26 & 4x61 & 120 & $88.4 \pm 20$\\
October 22 & 14:56:43.72 & 60605.6227282 & 5.48 & 4x61 & 115 & $89.9 \pm 21$\\
October 24 & 15:14:55.76 & 60607.6353676 & 6.71 & 4x61 & 82 & $94.4 \pm 28$\\
October 25 & 15:23:16.87 & 60608.6411675 & 7.33 & 4x61 & 30 & $9.2 \pm 68$\\
October 25 & 15:36:41.71 & 60608.6504828 & 7.33 & 4x61 & 63 & $35.7 \pm 48$\\
November 07 & 15:49:36.70 & 60621.6594525 & 15.29 & 4x61 & 101 & $45.4 \pm 24$ \\
November 08 & 15:48:49.39 & 60622.6589050 & 15.90 & 4x61 & 97 & $116.6 \pm 27$\\
November 09 & 12:56:10.62 & 60623.5390118 & 16.44 & 4x61 & 113 & $84.7 \pm 21$\\
November 10 & 14:36:39.34 & 60624.6087886 & 17.09 & 4x61 & 122 & $4.3 \pm 20$\\
November 11 & 13:34:48.52 & 60625.5658393 & 17.68 & 4x61 & 103 & $95.5 \pm 23$\\
November 12 & 12:41:44.47 & 60626.5289870 & 18.26 & 4x61 & 85 & $109.5 \pm 35$\\
November 15 & 13:46:35.32 & 60629.5740199 & 20.13 & 4x61 & 104 & $55.4 \pm 24$ \\
November 20 & 15:19:18.07 & 60634.6384036 & 23.22 & 4x61 & 97 & $27.6 \pm 25$\\
November 21 & 15:45:46.72 & 60635.6567907 & 23.84 & 4x61 & 107 & $140.9 \pm 23$\\
November 23 & 15:18:05.71 & 60637.6375661 & 25.06 & 4x61 & 91 & $85.3 \pm 31$\\
November 24 & 14:47:14.52 & 60638.6161402 & 25.65 & 4x61 & 68 & $83.8 \pm 44$\\
November 24 & 14:54:16.01 & 60638.6210186 & 25.66 & 4x61 & 55 & $17.1 \pm 66$\\
December 20 & 10:36:40.06 & 60664.4421303 & 41.44 & 4x61 & 97 & $54.5 \pm 26$\\
December 23 & 9:25:50.45 & 60667.3929450 & 43.25 & 4x61 & 111 & $99.2 \pm 21$\\
\hline
\multicolumn{7}{c}{ \textbf{PM J05408-3323}\rule{0pt}{2.5ex}} \\
\hline
November 09 & 12:25:38.07 & 60623.5178018 & 0.00 & 4x184 & 113 & $5.3 \pm 6$ \\
November 11 & 11:48:26.83 & 60625.4919772 & 0.12 & 4x153 & 101 & $19.7 \pm 6$\\
November 13 & 12:07:47.70 & 60627.5054132 & 0.24 & 4x150 & 101 & $1.9 \pm 7$\\
November 15 & 12:23:02.83 & 60629.5160050 & 0.36 & 4x176 & 101 & $-14.0 \pm 48$\\
November 20 & 11:03:37.58 & 60634.4608516 & 0.66 & 4x162 & 102 & $-1.9 \pm 6$\\
November 22 & 12:37:06.88 & 60636.5257741 & 0.79 & 4x171 & 101 & $-4.0 \pm 7$\\
November 24 & 12:16:15.41 & 60638.5112894 & 0.91 & 4x181 & 97 & $-3.4 \pm 7$\\
November 26 & 10:10:20.20 & 60640.4238449 & 1.02 & 4x174 & 102 & $-0.9 \pm 7$\\
December 04 & 11:33:45.58 & 60648.4817775 & 1.51 & 4x150 & 102 & $-1.3 \pm 6$\\
December 06 & 11:42:13.21 & 60650.4876529 & 1.63 & 4x184 & 80 & $-3.1 \pm 7$\\
December 20 & 9:07:40.66 & 60664.3803317 & 2.48 & 4x182 & 89 & $-16.8 \pm 8$\\
December 23 & 8:37:24.86 & 60667.3593155 & 2.66 & 4x156 & 101 & $-3.0 \pm 6$\\
\hline
\end{tabular}
\end{table}

\end{appendix}

\FloatBarrier 
\clearpage

\end{document}